\documentclass[a4paper,fleqn,usenatbib]{mnras}

\usepackage[T1]{fontenc}
\usepackage{ae,aecompl}
\usepackage{graphicx}	
\usepackage{graphics}
\usepackage{amsmath}	
\usepackage{amssymb}
\usepackage{float}

\usepackage{color}
\usepackage{soul}

\newcommand{\kms}{${\rm\,km\,s^{-1}}$} 

\title[{\it Gaia}-DR2 white dwarf population identification]{Random Forest identification of the thin disk, thick disk and halo {\it Gaia}-DR2 white dwarf population}

\author[S. Torres et al.]{
S. Torres$^{1,2}$\thanks{E-mail: santiago.torres@upc.es},  C. Cantero$^{1}$, A. Rebassa-Mansergas$^{1,2}$, G. Skorobogatov$^{1}$,
\newauthor 
F. M. Jim\'enez-Esteban$^{3,4}$, E. Solano$^{3,4}$   
\\
$^{1}$ Departament de F\'{\i}sica, Universitat Polit\'{e}cnica de Catalunya, c/Esteve Terrades 5, 08860 Castelldefels, Spain\\
$^{2}$ Institut d'Estudis Espacials de Catalunya, Ed. Nexus-201, c/Gran Capit\'a 2-4, 08034 Barcelona, Spain\\
$^{3}$Departmento de Astrof\'{\i}sica, Centro de Astrobiolog\'{\i}a (CSIC-INTA), ESAC Campus, Camino Bajo del Castillo s/n,\\
E-28692 Villanueva de la Ca\~nada, Madrid, Spain\\
$^{4}$Spanish Virtual Observatory, Spain\\
}

\date{Accepted XXX. Received YYY; in original form ZZZ}

\pubyear{}
\begin{document}
\label{firstpage}
\pagerange{\pageref{firstpage}--\pageref{lastpage}}
\maketitle

\begin{abstract}
{\it Gaia}-DR2 has provided an unprecedented number of white dwarf candidates of our Galaxy. In particular, it is estimated that {\it Gaia}-DR2 has observed nearly 400\,000 of these objects and close to 18\,000 up to 100 pc from the Sun. This large  quantity of data requires a thorough analysis in order to uncover their main Galactic population properties, in particular the thin and thick disk and halo components. Taking advantage of recent developments in artificial intelligence techniques, we make use of a detailed Random Forest algorithm to analyse an 8-dimensional space (equatorial coordinates, parallax, proper motion components and photometric magnitudes) of accurate data provided by {\it Gaia}-DR2 within 100\,pc from the Sun. With the aid of a thorough and robust population synthesis code we simulated the different components of the Galactic white dwarf population to optimize the information extracted from the algorithm for disentangling the different population components. The algorithm is first tested in a known simulated sample achieving an accuracy of 85.3\%. Our methodology is thoroughly compared to standard methods based on kinematic criteria demonstrating that our algorithm substantially improves previous approaches. Once trained, the algorithm is then applied to the {\it Gaia}-DR2 100 pc white dwarf sample, identifying  12\,227 thin disk, 1410 thick disk and 95 halo white dwarf candidates, which represent a proportion of 74:25:1, respectively. Hence, the  numerical spatial densities are  $(3.6\pm0.4)\times10^{-3}\,{\rm pc^{-3}}$, $(1.2\pm0.4)\times10^{-3}\,{\rm pc^{-3}}$ and $(4.8\pm0.4)\times10^{-5}\,{\rm pc^{-3}}$ for the thin disk, thick disk and halo components, respectively. The populations thus obtained represent the most complete and volume-limited samples to date of the different components of the Galactic white dwarf population.
\end{abstract}


\section{Introduction}
\label{s-intro}

White dwarfs are the most common evolutionary remnants among the stellar population. Considering a standard initial mass function, nearly 97\% of all main sequence stars in the Galaxy will finish or have already finished their lives as white dwarfs \citep[e.g.][]{Fontaine2001}.  Since nuclear fusion reactions have ceased in the interiors of white dwarfs, electron degeneracy pressure is the only mechanism working against gravitational collapse. White dwarfs are hence doomed to a long and slow cooling process in which potential gravitational energy is released through a thin atmosphere, in most cases composed of hydrogen and in a smaller number of cases of helium. From a theoretical point of view, we can reasonably state that the evolution of white dwarfs is relatively well understood \citep[see, for instance][]{Mestel1952,Althaus2010}. 

As long living objects, white dwarfs carry then important information not only about the evolution of their progenitors stars, but also on the properties of their parent populations. In this sense, white dwarfs have been widely used as reliable cosmochronometers, as well as to characterize the ensemble properties of several populations such as the Galactic thin and thick disk  \cite[e.g.][]{Winget1987,GBerro1988,GBerro1999,Torres2002,Rowell2011,Rowell2013}, the halo \cite[e.g.][]{Mochkovitch1990,Isern1998,GBerro2004,Cojocaru2015,Kilic2018}, and the bulge \citep{Calamida2014, Torres2018}, or to carry out precise studies of Galactic open and globular clusters \cite[e.g.][]{GBerro2010, Jeffery2011, Hansen2013, Torres2015}. Furthermore, using white dwarfs as tracers of the Galactic evolution has played a key role in understanding many capital problems. For instance, since the MACHO collaboration experiment for the microlensing detection \citep[e.g.][]{Alcock2000}, halo white dwarfs have been initially suggested as natural candidates to contribute to the dark matter content of the Galaxy \citep[e.g.][]{Oppenheimer2001},  although later studies demonstrated that this contribution is rather limited \citep[e.g.][]{Torres2002,Flynn2003,GBerro2004,Kilic2004,Bergeron2005}. In the same way, the structure and kinematics of the Galactic disk have been the focus of intense studies. In particular, the structure and kinematics of the thick disk \citep[e.g.][]{Chiba2000,Carollo2010} is intimately linked to the characterization of high-proper motion thick disk white dwarfs \citep[e.g.][]{Reid2005,Fuhrmann2012}. Hence, the motivation for disentangling the white dwarf Galactic components is multiple. 

It has to be emphasized however that this is not a straight forward task, as the identification of white dwarfs belonging to a certain Galactic component suffers from two major drawbacks. First, due to the large surface gravity acting in white dwarf atmospheres, there is a broadening of Balmer's spectral lines thus generally impeding accurate radial velocity determinations, unless medium or high-resolution spectra are available \citep{Pauli2006,Anguiano17}. Second, by the action of this same high surface gravity, the metal content on these atmospheres sinks well below the deep interior of these stars. Consequently, most white dwarfs are absent of accurate radial velocity measurements and of  metallicity estimates\footnote{Note however that the metallicities of the white dwarf progenitors can be obtained from the main sequence companions if the white dwarfs are in binary systems \citep{Rebassa2016}}. We may add to this non-optimistic situation the fact that many white dwarfs have been found by high proper motions or by photometric surveys targeting blue or UV excess objects. Consequently most white dwarfs are absent of parallax measurements too, belonging then to incomplete magnitude-limited samples. These drawbacks forced the traditional membership classification procedure  to be based generally on kinematic criteria under the assumption of a zero radial velocity (in some cases equivalently to a zero perpendicular Galactic velocity, $W=0$), an assumption that has not been exempt from controversy \citep[e.g.][]{Reid2001, Koopmans2001} due to the existing overlap between the kinematic properties of the different Galactic components. In fact, the kinematic criteria on its own do not guarantee a proper membership classification and a complementary analysis of the cooling time and age estimate of the white dwarf is also needed \citep[e.g.][]{Hansen2001,Bergeron2003}.

Nevertheless, the advent of large-scale  automated surveys such as the Sloan Digital  Sky Survey \citep{2000AJ....120.1579Y},  the Pan-STARRS collaboration \citep{2002SPIE.4836..154K}, the RAVE Survey \citep{2008AJ....136..421Z}, the LAMOST \citep{Zhao2012},   and   the   SuperCosmos Sky Survey \citep{1998MNRAS.298..897H}, have provided us with  an unprecedented  wealth  of information that has allowed deepening our knowledge of the different stellar populations of the Galaxy. However, building a complete and volume--limited sample of the white dwarf population has been restricted to no more than a few tens of parsecs; an $86\%$ completeness for the 20\,pc sample and no more than  $70\%$ for the extended version up to 25\,pc \citep{Holberg2008, Holberg2016}. The {\it Gaia} mission has recently improved notably this situation as it has provided us with an unprecedented amount of precise astrometric and photometric data for more than one billion sources \citep{GaiaDR12016,GaiaDR22018}. Within {\it Gaia} DR2, $\sim 260\,000$ high-confidence white dwarf candidates have been proposed \citep{Fusillo2018} and it has also been shown that {\it Gaia} DR2 provides a practically complete and volume-limited sample of white dwarfs up to 100\,pc \citep{Jimenez2018}.

The {\it Gaia} mission provides us with an $n$-dimensional space of high quality and large quantity of astrometric and photometric measurements which can be used to predict the white dwarf membership population. However, this fact demands more sophisticated classification strategies. In this era of big data mining, automated intelligent artificial algorithms based on machine learning techniques are required. These techniques are not new and have been widely used in many fields  of astrophysics. For instance, Neural Network algorithms have been successfully used to discriminate stars from galaxies \citep{Bazell1998}, to classify galaxies according to their morphology \citep{Naim1995}, or to study the fraction of binaries in star clusters, \citep{SerraRicart1996}. These techniques have also been successfully used to classify populations in the {\it Hipparcos} Input Catalogue \citep{Hernandez1994} or in the Villanova White Dwarf Catalogue \citep{Torres1998}.

Currently, additionally to Neural Network algorithms, there is an appreciable number of Machine Learning methods oriented to automatically classifying large datasets, such as Support Vector Machines \citep{Joachims1999}, Na\"{\i}ve Bayes \citep{Witten2005}, Random Forest \citep{Breiman2001}, Decision Trees \citep{Quinlan1986} and Bagged Trees, to cite a few examples. There is no universal criterion to select which is the most appropriate for addressing a particular problem. However, one of the most promising ones because of its suitable performance on classifying different kinds of datasets, its execution time, its simplicity when tuning the initial free--parameters and its flexibility is the Random Forest algorithm \citep{Caruana2006,Wainberg2016}. Successful applications of the Random Forest algorithm in Astronomy include periodic variable star classification from {\it Hipparcos} data  \citep{Dubath2011}, quasars, stars and galaxies discrimination \citep{Gao2009}, and star classification in the Galactic center \citep{Plewa2018}, to cite some representative examples.

In the present work we aim to disentangle the white dwarf population within 100\,pc recently observed by the {\it Gaia} DR2 by means of a Random Forest algorithm. In the learning process we take advantage of our detailed population synthesis code, which is able to accurately reproduce an 8-dimensional space of astrometric and photometric properties for the thin and thick disk and the stellar halo of the Galaxy. 

The paper is structured as follows. In Section 2 we provide a general description of the Random Forest algorithm. The synthetic population sample used for training the algorithm is explained in Section 3, while in Section 4 we describe the {\it Gaia} white dwarf sample that we aim to classify. In Section 5 we present a testbed of the algorithm providing some statistics and a thorough comparison with other methods traditionally applied for classifying the white dwarf sample. In Section 6 we apply our Random Forest algorithm to the 100\,pc {\it Gaia} observed sample and derive the main properties and parameters of the different identified Galactic components. Finally, we present our main conclusions in Section\,\ref{s:conclusions}.

\section{The classification method: the Random Forest Algorithm}
\label{s-RFA}

A Random Forest algorithm \citep{Breiman2001} is an Ensemble Machine  Learning method used for classification purposes. It belongs to the category of supervised methods, that is, an initial labeled sample (in which the class or group of each of the elements of the data is known) is needed to train the algorithm. Once the algorithm is trained, it is then applied to the sample that we aim at classifying. In what follows we explain the basics concepts of a Decision Tree and the Random Forest algorithm. A thorough explanation of the method and references therein can be found in \cite{Breiman2001} and some examples of Astronomy applications are in \cite{Gao2009}, \cite{Dubath2011}, and \cite{Plewa2018}.

\begin{figure*}
  \centerline{\includegraphics[trim=0mm 0mm 0mm 80mm,clip=true,width=1.4\columnwidth]{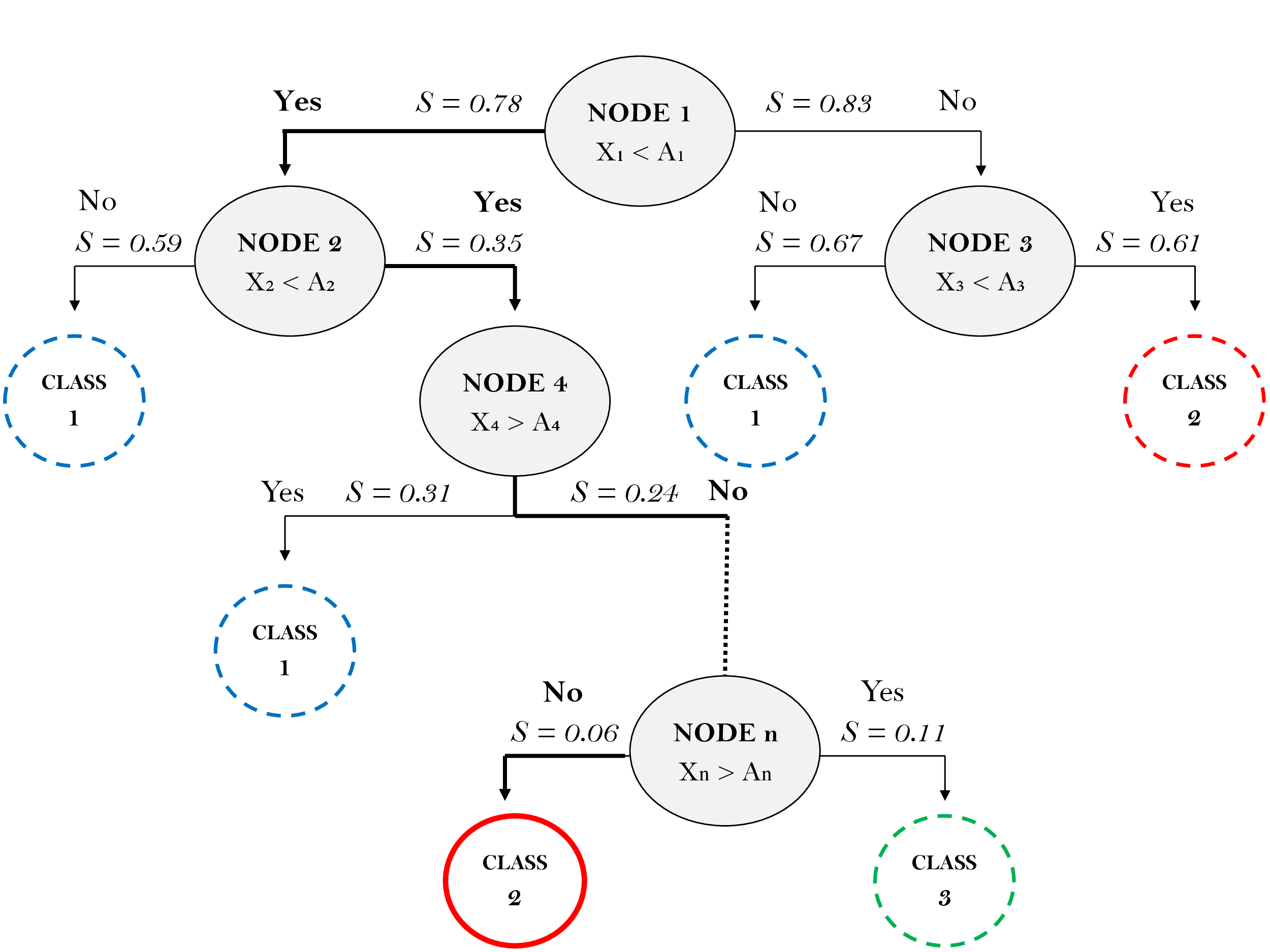}}
 \caption{Example of a Decision Tree structure with $n$-nodes and different levels of depth. The variables $X_{1}, X_{2}, \ldots , X_{n}$ on each node represent the features on the training sample, and $A_{1}, A_{2}, \ldots, A_{n}$ depict those values for splitting the data in the nodes. The entropy function $S$ is evaluated on each splitting branch (the values represented here are only for illustrative purposes). The branch that minimizes the entropy is marked as a bold black line. In this example, the star would be finally classified as class 2 out of three possible classes.}
 \label{root}
\end{figure*}

Random Forest algorithms are based on the principles of what is called the Decision Tree algorithm \citep{Quinlan1986}. The Decision Tree algorithm splits the initial labeled sample into two sub-samples: the training and the testing samples. As both samples are labeled, the former is used for training and constructing the model of classification and the latter is used as a testbed for evaluating the accuracy of that model. Through evaluating the training sample, the algorithm forms a rooted tree by means of different combinations of nodes in different depth levels. Each node splits again the training sample through numerical conditions depending on the variables of the sample. That is, each node basically evaluates a certain entropy function, in our case the Shannon entropy. According to the criteria used  to cut the training sample in the node, the entropy value will vary. Thus, the goal of the algorithm is to find the branching that minimizes the entropy and consequently classifies the data in the best possible way. An example of this process can be seen in Figure \ref{root}, which shows a small decision tree with $n$-nodes in several depth levels. Each node splits the sample comparing the features or variables of the training sample ($X_{1}, X_{2}, \ldots , X_{n}$) with certain values ($A_{1}, A_{2}, \ldots, A_{n}$). After the tree is built, it selects the path (bold line of Fig. \ref{root}) where the entropy, $S$, is minimum. In the example of Fig. \ref{root} the star would be finally classified as class 2 out of three possible classes.   

The Random Forest algorithm uses several of these decision trees (of the order of hundreds per each object of the sample)  for constructing the classification model.  This fact adds two extra-levels of randomness. First, the algorithm randomly selects a small portion of the training sample to construct each decision tree.  Therefore each tree is structured according to a different partition of the original training sample. Secondly, the features or variables used for splitting the data are randomly selected for each node of the different trees. Thus the algorithm combines many decision trees which are evaluated independently and the prediction will be the mean value (class) of all decision trees.  These extra degrees of randomness introduced by the Random Forest provide a larger flexibility of the algorithm. This results in greater tree diversity, which in turn permits to find the best features for classifying each object. As a consequence, the algorithm achieves a better overall classification and, at the same, avoids any possible over-fitting. Finally, once the algorithm is trained and tested, it can be applied to any sample that we want to classify, within the same variable space as the initial labeled sample.

As any Machine Learning technique, the implementation of the Random Forest algorithm requires the definition of some hyperparameters, i.e. free-parameters that the user can modify before the algorithm begins the learning process.  The most important ones are {\it max\_depth} and {\it estimators}. The former controls the number of levels in each decision tree and the latter controls the number of trees in the forest. In principle, depending on the configuration of these hyperparameters, the accuracy of the classification  can change. Therefore, some initial testing is needed in order to find the optimal set of these hyperparameters.  However, the robustness of the algorithm is high enough to guarantee that the final classification  remains unchanged for a wide range of hyperparameter values.

\section{The synthetic population sample}
\label{s:synt}

Our population synthesis code has been widely used in the study of the white dwarf population of the different Galactic components, as well as in globular and open clusters \citep[and references there in]{GBerro1999, GBerro2004, GBerro2010, Torres2001, Torres2002, Torres2015, Torres2016}. In particular, a comprehensive study of the capabilities of {\it Gaia} and previous estimates on the Galactic white dwarf population can be found  in \cite{Torres2015}, while a recent analysis of the {\it Gaia}-DR2 100 pc white dwarf population is presented in \cite{Jimenez2018}. In what follows we briefly describe the main ingredients employed in our simulations and point the reader to \cite{Jimenez2018} for a detailed description of the code and a full list of references.

The white dwarf population is simulated according to a three-component Galactic model  \citep[e.g.][]{Chiba2000,Castellani2002,Reid2005}: thin and thick disk, and halo or stellar spheroid. In particular, for the thin disc population, we adopt an age of 9.2\,Gyr \citep{Jimenez2018} with a constant star formation rate, while the spatial distribution of the synthetic stars follows a double exponential profile with a scale height of 250\,pc and a scale length of 2.6\,kpc. The thick disc is modeled from a star formation rate peaked at 10\,Gyr in the past and extended up to 12\,Gyr. Similarly, the thick disc population follows a double exponential spatial distribution  with a scale height of 1.5\,kpc and a scale length of 3.5\,kpc. The birth time for synthetic stars of the halo population is randomly assigned within a burst of constant star formation lasting 1\,Gyr that happened 13.5\,Gyr in the past. Besides, halo stars are located according to an isothermal distribution. Kinematic properties of each Galactic component are drawn according to the standard Schwarzschild ellipsoid with diagonal dispersion tensor in the usual $UVW$ Galactic coordinate system, with $U$  towards the Galactic center, $V$ in the direction of the plane rotation and $W$ perpendicular to the  Galactic plane pointing and positive towards the North Galactic Pole. The average values for the velocity components and their respective dispersions are the observed estimates by \cite{Rowell2011}. The peculiar velocity of the Sun with respect to the Local Standard of Rest (LSR) adopted is $(U_{\odot},\,V_{\odot},\,W_{\odot})=(7.90,\,11.73,\,7.39)\,$\kms \citep{Bobylev2017}. The mass of main-sequence stars are randomly chosen following a Salpeter-like initial mass function with $\alpha=-2.35$. 

Once we determine which stars have become white dwarfs, we evaluate their cooling evolution by means of an updated set of white dwarf evolutionary cooling sequences provided by La Plata Group -- see \cite{Althaus2015} and \cite{Camisassa2016,Camisassa2017} and references therein.  These sequences are metallicity dependent and encompass the full range of white dwarf masses for CO-core and ONe-core  white dwarfs. For the thin disk population we adopt a solar metallicity value, $Z=0.014$, a subsolar  $Z=0.001$ for the thick disk, and $Z=0.0001$ for halo stars. Additionally, for each of the white dwarfs generated, we consider hydrogen-rich or hydrogen-deficient atmospheres according to the canonical distribution of $80\%$  and $20\%$, respectively. Colours and magnitudes are then interpolated in the corresponding cooling sequences and calculated in {\it Gaia} filters (R. Rohrmann's private communication). The simulated synthetic population for the three Galactic components is then mixed proportionally to 75:20:5 for the thin disc, thick disc, and halo main-sequence stars, respectively. This ratio results in a proportion 77:17:6 for the thin disc, thick disc, and halo white dwarfs, respectively, in good agreement with the estimates by \cite{Rowell2011}. The simulated white dwarf sample is then normalized to the local space density of white dwarfs of $4.9\times 10^{-3}\,{\rm pc^{-3}}$ as estimated by \cite{Jimenez2018}. Finally, and in order to reproduce the observational uncertainties, we introduce a photometric and an astrometric error for each of our simulated objects based on {\it Gaia}'s performance\footnote{http://www.cosmos.esa.int/web/gaia/science-performance}. The final sample thus contains $\approx 18,000$ white dwarfs.

Finally, it is important to stress here that, as any supervised algorithm, the final classification depends on the initial prescriptions. For this reason, the characteristics of the Galactic populations adopted are as standard as possible in order to avoid any bias in the classification of the observed {\it Gaia} sample. For instance, a model with a too young thin disk age will automatically classify older objects as thick disk or halo candidates, regardless of their kinematic properties. Thus, we follow a strategy that first models the Galactic components in an overlapping space, and then trains the algorithm to disentangle each individual object. 

\section{The observed {\it Gaia}-DR2 100 pc white dwarf sample}
\label{s:obs}

Our population synthesis simulator permits us to establish the region of the Hertzsprung-Russell diagram where we expect to find the white dwarf population. This strategy has been used to identify  white dwarfs accessible by {\it Gaia} within 100 pc \citep{Jimenez2018}. The results derived in that work indicate that a distance of 100\,pc is the largest for which the {\it Gaia} white dwarf sample can be considered as both complete and volume-limited. Thus, we first delimit a wide region where all single white dwarfs can be found, regardless of the Galactic component they belong to. This region is defined by $M_{G}\,>\,2.95\times(G_{\rm BP}-G_{\rm RP})+10.83$ for $(G_{\rm BP}-G_{\rm RP})<1.2$ and $M_{G}\,>\,1.87\times(G_{\rm BP}-G_{\rm RP})+12.16$ for $(G_{\rm BP}-G_{\rm RP})>1.2$. Additionally, we apply the following criteria to the {\it Gaia} DR2 catalogue\footnote{http://gea.esac.esa.int/archive/}: 

\begin{itemize}
\item $\varpi>10\,$mas and $\varpi/\sigma_{\varpi}>10$
\item $F_{\rm BP}/\sigma_{F_{\rm BP}}>10$ and $F_{\rm RP}/\sigma_{F_{\rm RP}}>10$
\item ${\it phot\_bp\_rp\_excess\_factor}<1.3+0.06\times(G_{\rm BP}-G_{\rm RP})^2$ 
\item $H_G=G+5\log \mu+5>3.02\times(G_{\rm BP}-G_{\rm RP})+11.95$
\end{itemize}

The first two cuts select objects below 100\,pc with photometric and astrometric relative errors under $10\%$. The cut in {\it phot\_bp\_rp\_excess\_factor} prevents against photometric errors in the BP and RP bands, especially significant for faint sources in crowded areas \citep{Lindegren18}.  In a complementary way to this last cut, we  add a cut in reduced proper motion $H_G$ in order to mainly remove low proper motion contaminants. This limiting criterion is obtained by extrapolating the cooling track of a $0.6\,M_{\odot}$ DA white dwarf with a tangential velocity $V_{\rm tan}=5$\kms. The tangential speed adopted represents a trade-off between eliminating as many contaminants as possible and preserving the selected sample as complete as possible. 

\begin{figure}
  \centerline{\includegraphics[trim=5mm 30mm 5mm 35mm,clip,width=1.1\columnwidth]{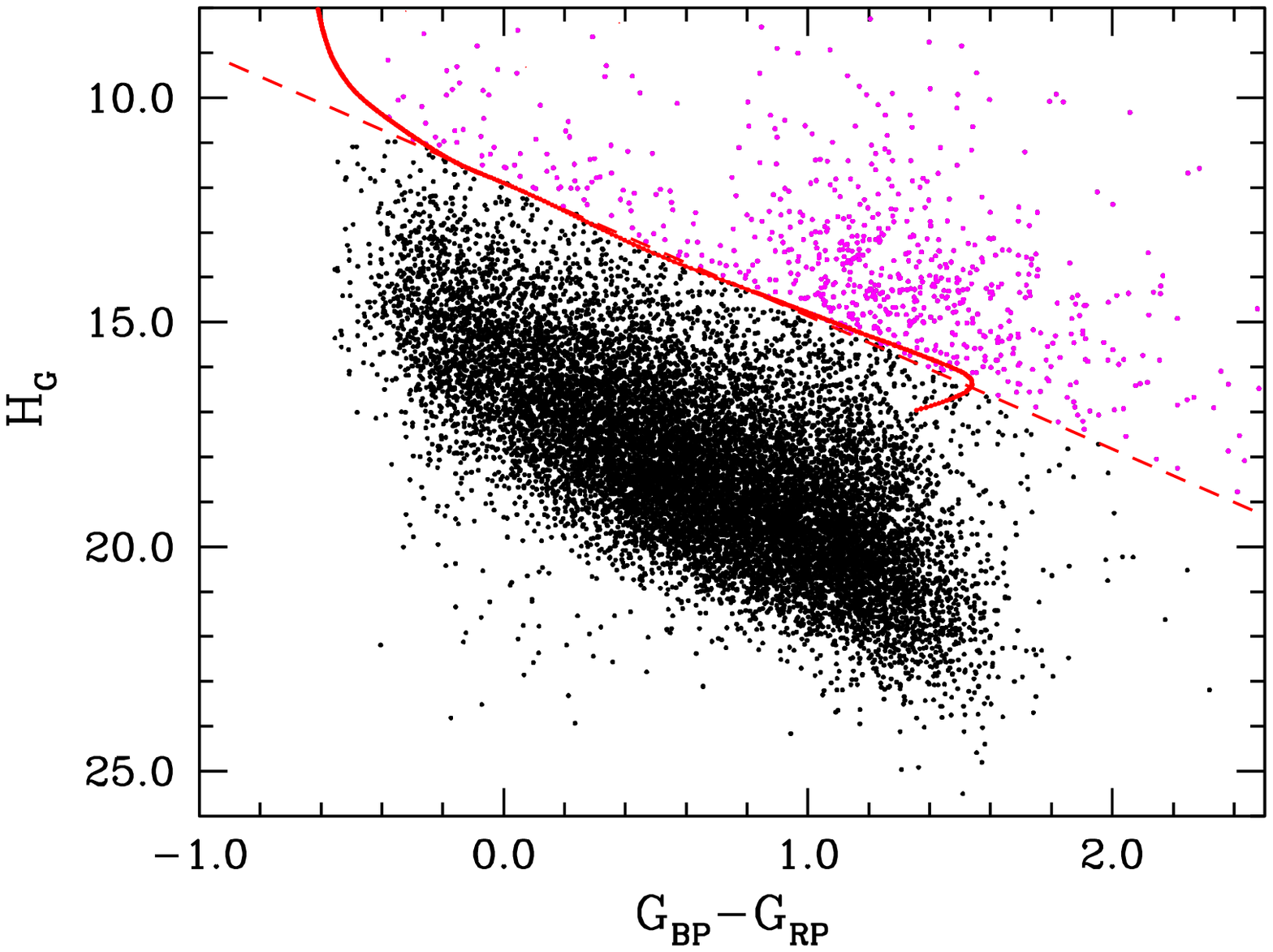}}
    \centerline{\includegraphics[trim=5mm 30mm 5mm 35mm,clip,width=1.1\columnwidth]{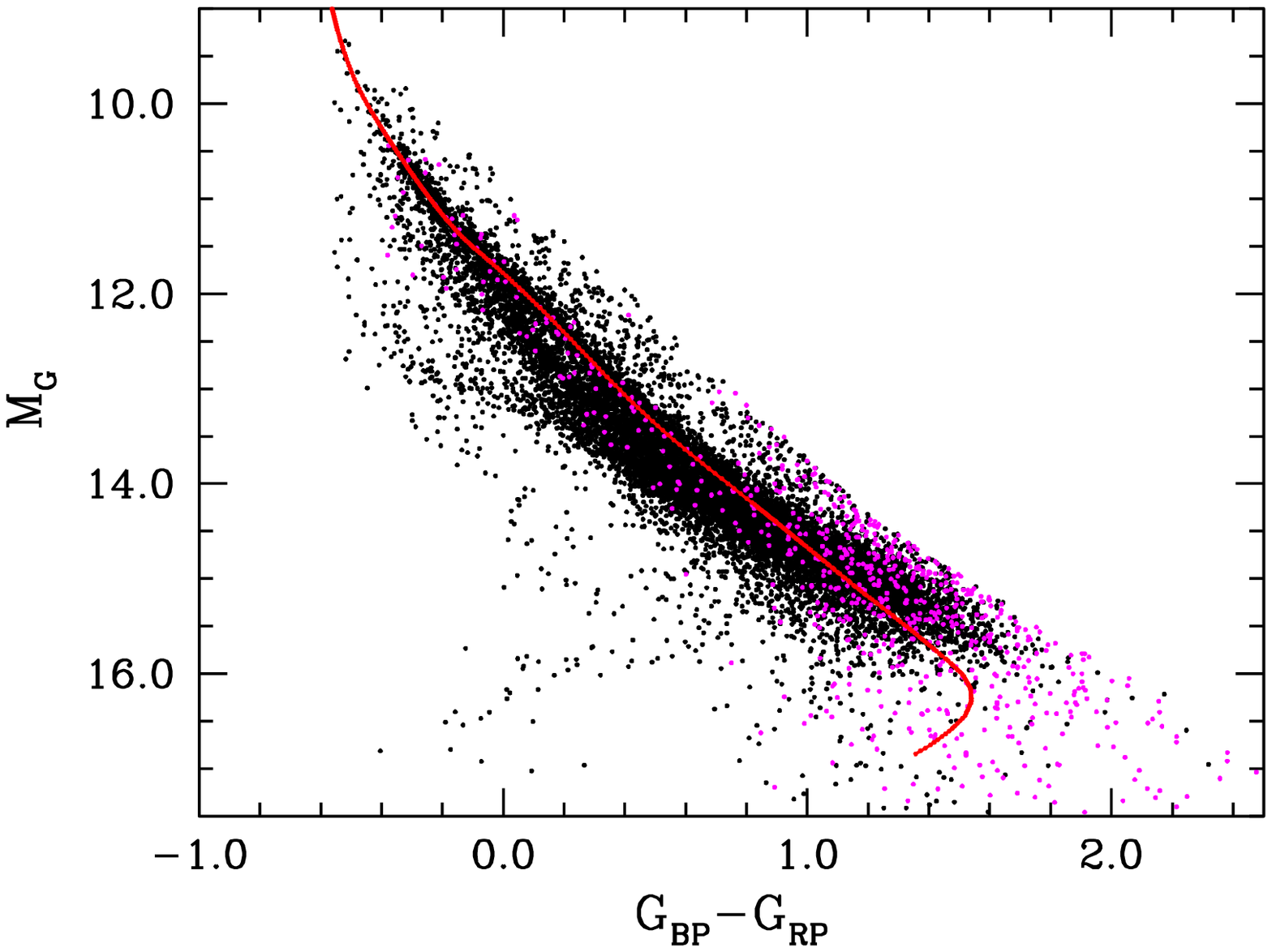}}
  \caption{Reduced proper motion diagram (top panel) for the white dwarfs accessible by {\it Gaia} within 100 pc. Also plotted is the cooling track for a $0.6\,M_{\odot}$ DA white dwarf \citep{Camisassa2016} with a tangential velocity $V_{\rm tan}=5$\kms (red solid line) and its extrapolated line (red dashed line). Objects above this line are eliminated (magenta dots). The corresponding Hertzsprung-Russell diagram for the selected objects (black dots) and those eliminated (magenta dots) is presented in the bottom panel.}
   \label{f:gaiasamples}
\end{figure}

In Figure \ref{f:gaiasamples} we plot the reduced proper motion diagram (top panel) for the white dwarfs accessible by {\it Gaia} within 100 pc. Also plotted is the cooling track for a $0.6\,M_{\odot}$ DA white dwarf \citep{Camisassa2017} with a tangential velocity $V_{\rm tan}=5$\kms (red solid line) and its extrapolated line (red dashed line). The Hertzsprung-Russel diagram for the selected objects is shown in the bottom panel of the same figure.  The eliminated objects by the reduced proper motion cut (magenta dots) represent $\sim8\%$ of the whole sample and are mainly faint red ($G_{\rm BP}-G_{\rm RP}>1$) (see the bottom panel of Fig. \ref{f:gaiasamples}). It is worth saying that in our analysis of the synthetic sample (see Fig. \ref{f:hredsim}) the expected percentage of eliminated objects is only $2\%$ and that these targets are not particularly red. Thus, similarly to the ${\it phot\_bp\_rp\_excess\_factor}$ cut, the reduced proper motion cut helps to remove faint sources in crowded areas, specifically those with low motion. 

The final selected sample contains a total of 13\,732 white dwarf candidates and represents $\sim60-75\%$ of the total population of white dwarfs within 100\,pc, assuming an estimate for the spatial density of $(4.9\pm0.4)\times10^{-3}\,{\rm pc^{-3}}$ \citep{Jimenez2018}. It is worth mentioning  that the {\it Gaia} sample under study in this work is an extension of the 100\,pc sample we presented in \cite{Jimenez2018}. This last sample contains 8343 CO-core and 212 ONe-core white dwarf candidates within the range of effective temperatures of 5000\,K$<T_{\rm eff}<$80\,000\,K together with a detailed analysis of their stellar parameters (luminosity, effective temperature, surface gravity and mass) that can be gathered from {\em The SVO archive of {\it Gaia} white dwarfs} at the Spanish Virtual Observatory portal\footnote{http://svo2.cab.inta-csic.es/vocats/v2/wdw}. The \cite{Jimenez2018} analysed sample does not include white dwarfs below 5000\,K. In the sample adopted here we do include these cooler white dwarfs since the effective temperature is not used by our Random Forest algorithm for the classification.

\section{Training and testing the algorithm}

In this section we present the results of our training and testing Random Forest algorithm applied to the synthetic white dwarf population sample described in section \ref{s:synt}. Additionally, we compare the results with other procedures frequently used in the literature to disentangle the Galactic components. It is worth noting here that for a proper application of the Random Forest algorithm a complete and statistically significant sample for each of the populations to classify is required. At present, there is no observed sample of the white dwarf population that meets these requirements. Hence, the use of a synthetic sample for the training and testing of the algorithm.

\subsection{Learning process}
\label{s-Super}

As previously explained in Section \ref{s-RFA}, our Random Forest algorithm is trained in a known sample, in this case provided by our population synthesis models. A search of the optimal hyperparameters used by the Random Forest model is firstly performed.  For our initial labeled sample of $\approx18\,000$ objects, a grid analysis of hyperparameter values reveals that a number of 500 decision trees per object ({\it estimators}$=500$) and a maximum number of 10 levels ({\it max\_depth}$=10$) per tree, provide an optimal performance of the algorithm. For larger values of the hyperparameters the error is only marginally reduced at expenses of a larger computational cost.

During the learning process the initial labeled sample is split into  training and testing samples. The latter provides a testbed for the validation of the classification model. In order to avoid over-fitting effects and other statistical artifacts, we apply the well-known {\textit{cross-validation}} technique\citep{Stone1974}. Basically, this technique randomly splits the initial sample into \textit{k}-folds (subsets of the data) which are iteratively used as training and testing samples. By doing this all the data of the initial sample is used for both training and validation. Finally,  the \textit{k}-folds results are averaged to produce a single estimation that maximizes the performance of the classification algorithm.

\subsubsection{Feature importance}

The variables initially used during the learning process,  provided by our population synthesis model, are those which are also available in the {\it Gaia}-DR2 observed sample. In particular, we have used the equatorial coordinates, $\alpha$ and $\delta$, the proper motion components, $\mu^*_{\alpha}\equiv\mu_{\alpha}\cos\delta$ and $\mu_{\delta}$, the parallax, $\varpi$, and the $G$, $G_{\rm BP}$, and $G_{\rm RP}$ pass-bands magnitudes. This set conforms an 8-dimensional space which provides a mix of spatial, astrometric, kinematic, and photometric properties of the white dwarf population.

The {\it importance} of a feature or variable indicates how useful this feature is in the construction of the decision trees of the classification algorithm. The $importance$ is calculated for a single decision tree by the amount that each attribute split point improves the performance measure, weighted by the number of observations the node is responsible for. The feature $importances$ are then averaged across all the Decision Trees within the Random Forest. In short, the more an attribute is used to make a key decisions within decision trees, the higher its relative $importance$. Finally, this $importance$ is calculated explicitly for each variable of the sample, allowing variables to be ranked and, eventually, discarded due to their lack of relevance.

\begin{figure}
  \centerline{\includegraphics[trim=5mm 30mm 5mm 35mm,clip,width=1\columnwidth]{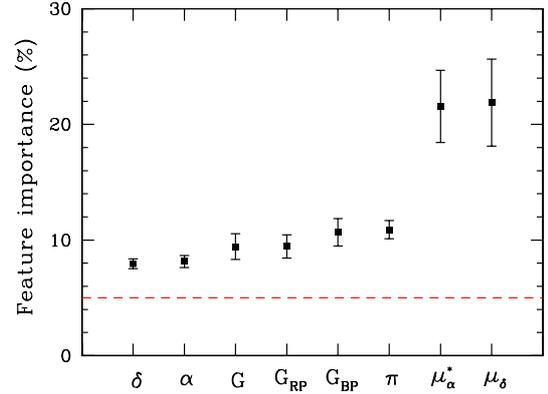}}
  \caption{Feature $importance$ for each of the variables used in the training sample. A significance threshold of a $5\%$ is marked as a red dashed line. See the text for details.} 
  \label{f:features}
\end{figure}

In Figure \ref{f:features} we show the $importance$ percentage for each of the variables considered by our Random Forest algorithm. The error bars give us an idea of how these $importances$ could vary when the training process is repeated with a different subset. 
In view of Fig.\ref{f:features}, it is clear that the proper motion components are the dominant features during the classification process. Together, both components represent nearly $44\%$ of the decisions when classifying white dwarfs stars. The parallax and the band-passes have approximately the same $importance$ of $\sim$10\,\%, while equatorial coordinates seem to be the less significant features in the classification process, with a relative $importance$ of $\sim8\%$ for each coordinate. It is also worth emphasizing here that the use of other variables that can be constructed as combinations of the previous ones, e.g. total proper motion, reduced proper motion, tangential velocity, etc., would not contribute with more information to the training process (see \ref{s:compa} for a comparison with other methods). For instance, we may substitute the equatorial proper motion and the $G$ magnitude by the reduced proper motion $H_G$ (see eq. \ref{e:hg}). The resulting importance for this new feature would be $\approx 40\%$. However, the global performance of the algorithm is reduced, given that the information contained in the individual variables is lost.

In this sense, the $importance$ analysis also reveals another relevant characteristic of the training process. Although it depends on the number of features and the typology of the data, an $importance$ above  $\sim1-5\%$ \cite[e.g.][]{Rafi2018} is usually adopted as a threshold limit for considering a variable as significant. Even when adopting a conservative $5\%$ (marked as red dashed line on Fig. \ref{f:features}), all the variables used in our training process clearly have a relevant $importance$. In other words, the algorithm is using the information contained in all the variables to a greater or a lesser extent. The kinematic properties are the dominant in the classification process, as logically expected for kinematically distinct populations. It has to be stressed however, that the information contained in the photometry is also relevant given that it is related to the different ages and cooling of the Galactic white dwarf components. In the same way, the spatial distribution of the disk and the halo populations is expected to follow different patterns that should ultimately be reflected in the equatorial coordinate distributions. Consequently, we consider all the features as relevant in the classification process.

\subsubsection{Confusion matrix analysis of the Random Forest algorithm}
\label{ss:confurf}

The confusion matrix is a well-used tool to evaluate the performance of  a classification algorithm, in particular for supervised algorithms such as the Random Forest. It compares the true values (rows) with the predicted values (columns) for the different groups of the classification. The indexes applied in our case are $i=1,2,3$ for the thin disk, the thick disk and the halo, respectively. Thus, $C_{11}$ stands for the number of actual thin disk stars classified as thin disk, $C_{12}$ as the number of actual thin disk stars classified as thick disk stars, $C_{13}$ as the number of actual thin disk stars classified as halo stars, and so on. To simplify the resulting figures of the confusion matrix we divide each term by the total number of objects of our simulation, which is of the same order as the expected number of white dwarfs within 100\,pc, i.e. $\approx$ 18\,000 objects. For an ideal classification, where all the elements are well-identified (with its class), the confusion matrix should appears as: 
\begin{equation}
C_{\rm ideal}=\left(
\begin{array}{ccc}
   0.77  & 0 & 0\\
    0 & 0.17 & 0 \\
    0 & 0 & 0.06
  \end{array}
  \right)
\end{equation}
reflecting the relative proportions of 77:17:6 for the thin disk, thick disk, and halo white dwarfs, respectively, of our synthetic sample.

The resulting confusion matrix obtained by our Random Forest algorithm applied to our synthetic population sample is as follows:

\begin{equation}
\label{crf}
C_{\rm RF}=\left(
\begin{array}{ccc}
   0.753  & 0.020 & 0\\
    0.113 & 0.053 & 0.002 \\
    0.003 & 0.009 & 0.047
  \end{array}
  \right)
\end{equation}

A bunch of valuable information can be extracted from this matrix. First of all, the accuracy of the method, defined as $Acc_{\rm RF}=\mathrm{Tr}(C_{\rm RF})$, represents the probability that an object of the sample is correctly classified. In our case $Acc_{\rm RF}=0.853$ or, in other words, $85.3\%$ of the white dwarfs are well classified. Secondly, if we look more specifically to each of the Galactic components, we see that $97\%$ of the thin disk white dwarfs are correctly classified\footnote{The probability, $P_k$, that the population $k$ is correctly classified is $P_k=\frac{C_{kk}}{C_{k1}+C_{k2}+C_{k3}}$.}. Halo white dwarfs are also reasonably well classified, with $80\%$ of these objects properly identified. However, the algorithm has only been able to identify $32\%$ of the thick disk stars, being the major part of these stars misclassified as thin disk white dwarfs. 

Finally, we can also estimate the contamination\footnote{The contamination of population $k$, $C_k$, is determined as $C_k=\frac{\sum_{i\ne k}C_{ik}}{C_{1k}+C_{2k}+C_{3k}}$} of a certain Galactic component. Objects classified as thin disk white dwarfs result in a clean sample, affected only by a $13\%$ level of contamination. The contamination of the group of thick disk stars is slightly higher, $35\%$. A specially clean sample results for the halo component, only contaminated by a small $4\%$ of thick disk white dwarfs.

In view of these results we conclude that our Random Forest algorithm succeeds at identifying clean and low-contaminated white dwarf populations belonging to the thin disk and the halo population. However, thick disk white dwarfs are only moderately identified due to their greater overlap with the thin disk population.

\subsection{Comparison with other classification methods}
\label{s:compa}

In this section we assess the ability of other methods frequently applied in the literature to disentangle the Galactic white dwarf components and compare the results to those obtained by our Random Forest algorithm. Given the lack of a spectroscopic analysis for a substantial fraction of white dwarfs, and the impossibility to accurately determine the metallicity of individual objects, the criteria frequently applied rely on the kinematic properties of the sample.

\subsubsection{Tangential velocity cut}

The tangential velocity can be used as a criterion for Galactic component classification in proper motion surveys for which trigonometric parallaxes, or at least photometric parallaxes, are available, joint to the absence of radial velocity measurements. Usually, a tangential velocity limiting cut in the range of $150>V_{\rm tan}>250\,$\kms\ is applied to distinguish among stars belonging to an spheroidal non-rotating population to those of the rotating disk. Additionally, a low tangential velocity limiting cut is also included in order to eliminate low motion contaminants. However, this criteria does not avoid the overlap between the different Galactic components in the tangential velocity distribution \citep[see, for instance, Fig. 16 from][]{Rowell2011}. 

We estimate the accuracy of this method by deriving the corresponding confusion matrix, $C_{\rm V_{tan}}$. We adopt a limiting cut of $20<V_{\rm tan}<90$\kms\ to identify thin disk stars, $90\le V_{\rm tan}\le 200$\kms\ for thick disk, and $V_{\rm tan}>200$\kms\ for halo stars. The resulting confusion matrix is:

\begin{equation}
C_{\rm V_{tan}}=\left(
\begin{array}{ccc}
   0.720  & 0.022 & 0\\
    0.128 & 0.059 & <0.001 \\
    0.005 & 0.019 & 0.047
  \end{array}
  \right)
\end{equation}

The accuracy of this procedure is $Acc_{\rm V_{tan}}=0.826$, which is smaller than the Random Forest accuracy. Note that the inclusion of a low tangential velocity cut slightly changes the proportion of the three components, specifically, the percentage of thin disk stars is reduced. The percentage of thin and thick disk white dwarfs properly classified is $97\%$ and $32\%$, respectively, quite similar to the Random Forest result. However, halo white dwarfs are only identified in $66\%$ of the cases. The estimated contamination of this last group is lower than $<1\%$, a value that is achieved at the cost of loosing a big portion of the slowly moving objects of the halo population. On the other hand, the contamination of the disk groups has increased to $16\%$ for the thin disk and to a considerable high value of $41\%$ for the thick disk population. It is worth mentioning that varying the limits of the tangential velocity values will change the probability of identification at the expense of a larger contamination, keeping the accuracy of the overall procedure more or less the same.

\begin{figure}
  \centerline{\includegraphics[trim=5mm 30mm 5mm 35mm,clip,width=1.1\columnwidth]{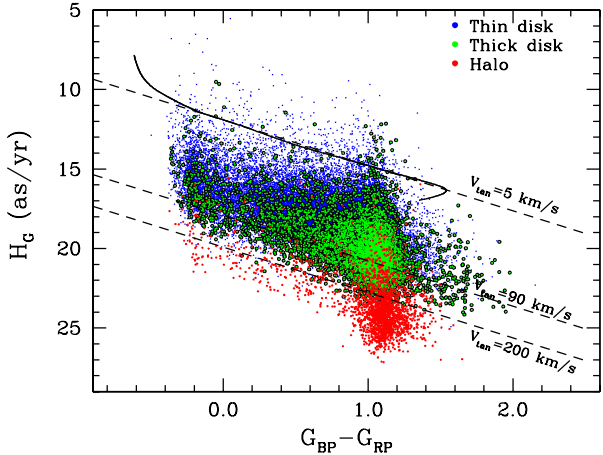}}
  \caption{Reduced proper motion diagram for our three Galactic components simulated sample. Also shown the cooling track for a $0.6\,M_{\odot}$ DA white dwarf \citep{Camisassa2016} with a tangential velocity $V_{\rm tan}=5$\kms (black solid line) and the corresponding extrapolated lines (black dashed lines) with $V_{\rm tan}=5, 90,   200$\kms\ for identifying the different Galactic populations.} 
  \label{f:hredsim}
\end{figure}

\subsubsection{Reduced proper motion diagram}
The reduced proper motion, defined as
\begin{equation}
\label{e:hg}
H_m=m+5\log \mu+5=M+5\log V_{\rm tan}-3.38,
\end{equation}
represents a pseudo-absolute magnitude that has been widely used as a selection criterion for disentangling populations with different kinematic characteristics \citep[e.g.][]{Kalirai2004,Kilic2006,Harris2006,Jimenez-Esteban11,Jimenez-Esteban12}. The reduced proper motion diagram, i.e. the reduced proper motion as a function of a given colour, is thus a widely used tool when parallax estimates are not available and white dwarfs cannot easily be identified in the Hertzsprung-Russell diagram. It allows to  disentangle in a reasonably efficient way the low-speed white dwarf population from cool main-sequence stars or, in particular, for high-speed halo subdwarfs. However it is less efficient at differentiating between the Galactic components of the white dwarf population. For instance, the highest-speed disk white dwarfs share the locus in the reduced proper motion diagram with the lowest-speed halo white dwarfs. 

In Figure \ref{f:hredsim}, we show the reduced proper motion $H_G$ as a function of the $G_{BP}-G_{RP}$ colour for the three Galactic components of our synthetic sample (blue, green and red dots for thin disk, thick disk and halo stars, respectively). Adopting the cooling track for a $0.6\,M_{\odot}$ DA white dwarf \citep{Camisassa2016} with a tangential velocity $V_{\rm tan}=5$\kms as a reference (black solid line), we determine the limiting line (black dashed line) for identifying the different populations. Thus, thin disk white dwarfs are those between the lines with  $5<V_{\rm tan}<90$\kms, thick disk stars those between the lines with  $90\le V_{\rm tan}\le 200$\kms  and $V_{\rm tan}>200$\kms for halo stars.

The resulting confusion matrix using this procedure is
\begin{equation}
C_{\rm H_G}=\left(
\begin{array}{ccc}
   0.693  & 0.062 & 0.003\\
    0.117 & 0.056 & 0.004 \\
    0.005 & 0.008 & 0.053
  \end{array}
  \right)
\end{equation}

The probabilities to correctly classify an object belonging to the thin disk, thick disk or halo components are $91\%$, $32\%$ and $80\%$, respectively, quite similar to that of the Random Forest algorithm. However, the contamination percentages, $15\%$, $56\%$ and $12\%$, respectively, are larger than those estimated from our machine learning algorithm. Specially significant is the percentage of misclassified objects in the thick disk group which, in the case of the reduced proper motion criteria, is more than half of the population. As in the tangential velocity cut method, the inclusion of a low limiting cut reduced the proportion of thin disk stars, slightly changing the final proportions. The accuracy of the reduced proper motion method is $Acc_{\rm H_{G}}=0.801$, smaller than in previous cases.

\subsubsection{Toomre diagram}

Finally, we analyze here another way to disentangle the Galactic components, based on the well known Toomre diagram \citep[e.g][]{Schronrich2009, Hawkins2015}. In this diagram the Galactic velocity perpendicular to the rotational direction is represented as a function of the rotational speed. We employ the usual criterion for the Galactic velocity components, $U$ towards the Galactic center, $V$ in the direction of the plane rotation and $W$ perpendicular to the  Galactic plane pointing and positive towards the North Galactic Pole. $UVW$ Galactic Velocities are expressed with respect to the Local Standard of Rest.

\begin{figure}
  \centerline{\includegraphics[trim=5mm 30mm 5mm 35mm,clip,width=1.1\columnwidth]{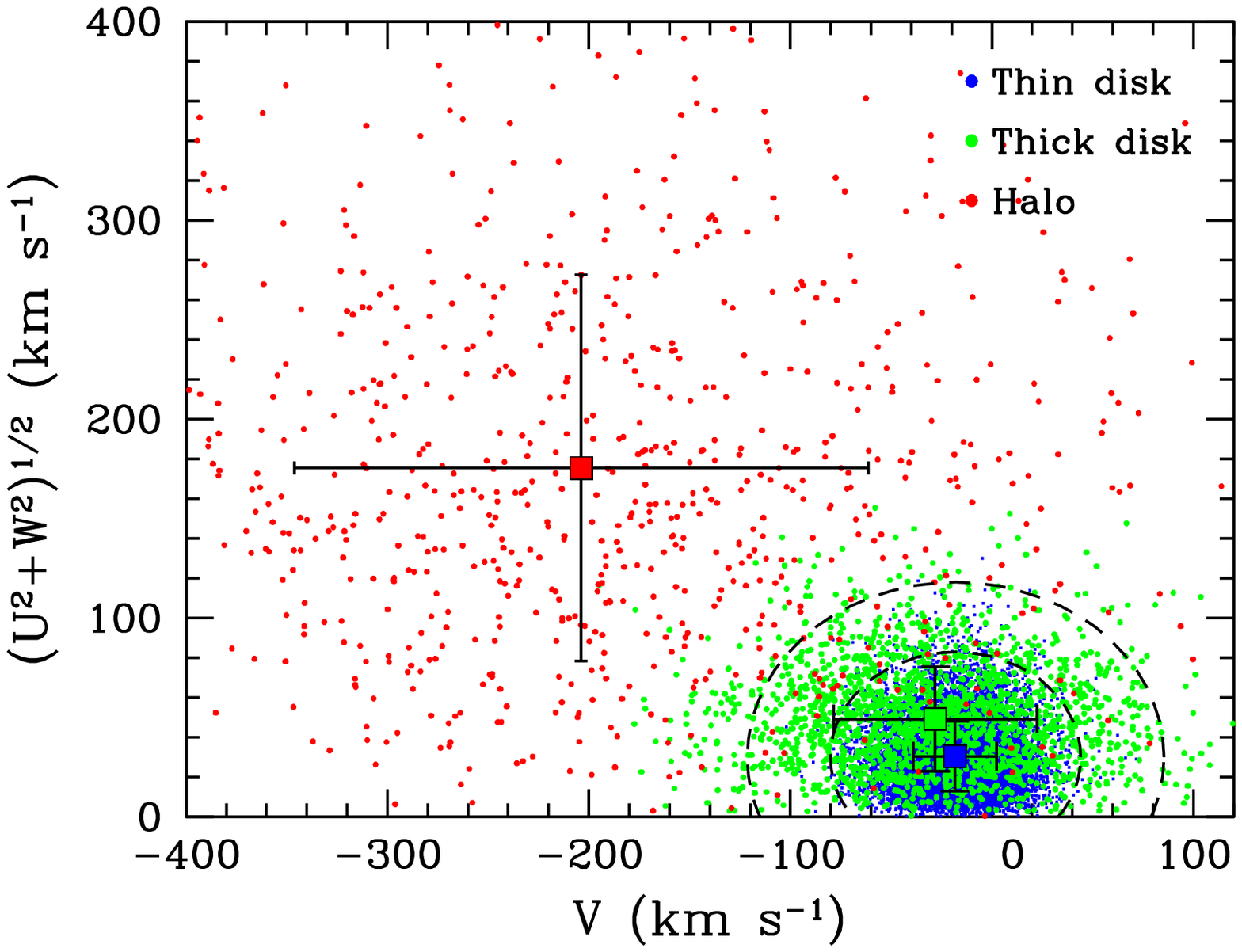}}
    \centerline{\includegraphics[trim=5mm 30mm 5mm 35mm,clip,width=1.1\columnwidth]{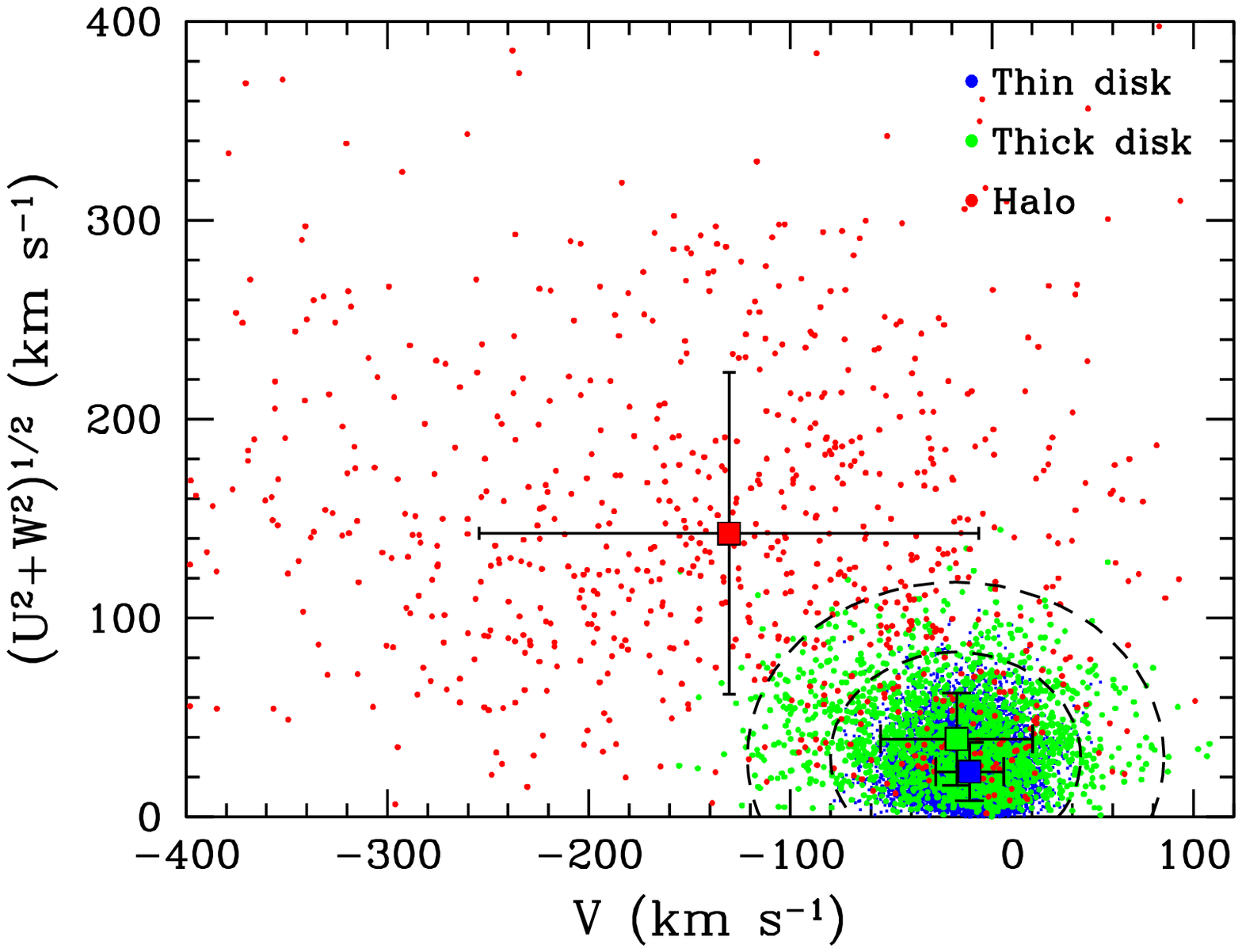}}
  \caption{Top panel: Toomre diagram for a complete 3-dimensional space velocity for our synthetic white dwarf sample. Bottom panel: the same but assuming a null radial velocity. The big squares represents the centroid of each Galactic component, and the vertical and horizontal lines their 1$\sigma$ error bars. Also plotted as dashed lines the  $3\sigma$ and $5\sigma$ thin disk contours. See text for details. }
  \label{f:toomres}
\end{figure}

In the top panel of Figure \ref{f:toomres} we show the Toomre diagram for our three-component synthetic Galactic population projecting the full 3-dimensional kinematic space in the $UVW$ Galactic velocity system. It is evident  that, although there is a clear overlap between the disk populations (blue and green dots for the thin and thick disk, respectively), the non-rotating component of the halo (red dots) seems to lay away from that of the rotating disk component and would be eventually identifiable. However, the main drawback of this procedure resides in the fact that the full 3-velocity space components are required to be known, a difficult task if no spectra are available from which one can determine the radial velocities. The usual procedure hence consists of projecting the proper motion components assuming a zero radial velocity \citep[e.g][]{Dehnen1998}. However, this fact is not exempt of a moduli reduction of the velocity components \citep{Pauli2006}. To exemplify this effect we have plotted the Toomre diagram of the different synthetic Galactic components in the bottom panel of Fig. \ref{f:toomres}, under the  assumption of zero radial velocity. To help with the visualization, we plot the centroid (big square) for each Galactic component along with its 1$\sigma$ error bars. It is clear that the null radial velocity assumption implies a reduction of the velocity components, especially significant for the halo population. The mean average moduli reduction for the thin disk, thick disk and halo components are $\Delta|V|=8.3,\ 11.4$ and $74\,$km/s, respectively. In particular, the mean for the halo Galactic components when the full velocity space is considered are $(U,V,W)=(-16.9,\, -203,\,-5.9)\,$km/s, while the resulting mean values when a zero radial velocity is assumed are $(U,V,W)=(-6.9,\,-130,\,-1.8 )\,$km/s. This fact implies that the efficiency to disentangle the Galactic populations by using the Toomre diagram is substantially restricted unless the full velocity space is used.

 Following a standard criteria based on the velocity ellipsoid \citep[e.g][]{Venn2004},  we select as halo members those white dwarfs whose velocities in the Toomre diagram depart $5\sigma$ from a typical disk distribution, while we consider as thick disk stars those who are in the interval between $3\sigma$ and $5\sigma$. Adopting this criteria we evaluate the confusion matrix for our synthetic population,

\begin{equation}
C_{\rm Toomre}=\left(
\begin{array}{ccc}
   0.766  & 0.007 & 0\\
    0.127 & 0.039 & 0.002 \\
    0.008 & 0.011 & 0.039
  \end{array}
  \right)
\end{equation}

\begin{table*}
\caption{Summary of the results obtained by the classification methods studied in this work. Columns 2 to 4 show the probability of each method to identify a white dwarf belonging to a particular group, while columns 5 to 7 represent the contamination of each group. The accuracies of each method, $Acc$, the $G-mean$, and the Cohen's $\kappa$ coefficients are presented in the last three columns, respectively.} 
\begin{center}
\begin{tabular}{cccccccccc}  
\hline \hline 
(1) & (2) & (3) & (4) & (5) & (6) & (7) & (8) & (9) & (10) \\
Method &  \multicolumn{3}{c}{\% Identification} & \multicolumn{3}{c}{\% Contamination} & $Acc$ & $G-mean$ & $\kappa$ \\
 & thin & thick & halo & thin & thick & halo & &  &  \\
\hline \hline

RF & 97 & 33 & 80 & 13 & 35 &  4 & 0.853 & 0.635 & 0.528\\ 
$V_{\rm tan}$ & 97 & 32 & 66 & 16 &  41 & <1 & 0.826 & 0.589 & 0.495 \\ 
$H_G$ & 91 & 32 & 80 & 15 & 56 & 12 & 0.801 & 0.620 & 0.444 \\  
Toomre & 99 & 23 & 67 & 15 & 32 & 5 & 0.844 & 0.536 & 0.465 \\ 
\hline \hline
\end{tabular}
\end{center}
\label{t:methods}
\end{table*}

The probability to identify thin disk stars, $99\%$, is extremely high, although the contamination presented for this group is moderate, $15\%$. However, the ability to identify thick disk white dwarfs is the lowest among all methods analyzed in this work, 
$23\%$, with a moderate contamination of $32\%$. With respect to halo white dwarfs, the Toomre diagram criterion only permits to recover $67\%$ of the spheroid population, although the contamination in the sample is very low, $5\%$. It is worth mentioning here that changing the $5\sigma$ and $3\sigma$ criteria will change the capability of the method to identify each of the Galactic components at the expense of varying the contamination in each of the groups. In view of the existing overlap presented by the different Galactic components (see Fig. \ref{f:toomres}), a decrease, for instance, of the $5\sigma$ limit for the halo population will increase the percentage of halo identified white dwarfs but, at the same time, it will increase the number of disk stars misclassified as halo stars and, consequently, the contamination of this group will also increase.

\subsubsection{Comparative summary}

We finish this section by summarizing the results obtained by the four classification methods analyzed in this work: Random Forest algorithm (RF), tangential velocity criterion ($V_{\rm tan}$), reduced proper motion diagram criterion ($H_G$) and Toomre diagram (Toomre). In Table \ref{t:methods} we show: in columns 2 to 4 the probability of each method to identify a white dwarf belonging to a particular group, in columns 5 to 7 the contamination of each group, the accuracy of each method in column 8, and in the last two columns two global estimators of the performances of each method (see the definitions and a discussion below).

In general terms, all the classification procedures succeed in identifying the thin disk population (with an identification percentage between $91\sim99\%$), while the halo population is moderately identified ($66\sim 80\%$). The thick disk population is hardly identified by any of the methods ($<33\%$). In this sense, the Toomre diagram criterion achieves the poorest result, leading us to discard this method for the thick disk identification. 

The degree of identification and the percentage of contamination can vary significantly from one method to another. On average, the highest percentages of identification are achieved by the Random Forest algorithm and the reduced proper motion criterion, while the lowest contamination percentages are obtained by the Random Forest and the Toomre diagram criterion.

A way to assess the performance of a classifier is by means of  a global estimator or measure \citep[e.g.][]{Labatut2011,Tharwat2018}. We recall that the accuracy, $Acc$, (see definition in Sec. \ref{ss:confurf}) represents the probability to identify a star belonging to a certain group. The Random Forest algorithm presents the highest accuracy, $Acc_{\rm RF}=0.853$, while the lowest corresponds to the reduced proper motion criterion, $Acc_{\rm H_{G}}=0.801$. The difference, not too large, between the accuracy of the methods reflects that they all succeed in classifying the thin disk population, which contains the majority of white dwarfs. An estimator that is independent of the relative proportions of the populations would be desirable. Among the many different estimators defined in the literature, we will consider two of them as illustrative examples. The $G-mean$\footnote{$G-mean=\left(\prod_{i=1}^3 P_i\right)^{1/3}$, where $P_i$ is the probability to identify objects of class $i$.} represents the geometrical mean of the probabilities of identification and it is independent of the ratio between the different populations. The second one is the well known Cohen's $\kappa$\footnote{The Cohen's $\kappa$ coefficient is defined as $\kappa=\frac{P_a-P_{\epsilon}}{1-P_{\epsilon}}$, where $P_a$ is the probability of agreement and $P_{\epsilon}$ is the probability of agreement due to chance. In our case, $P_a=Acc$ and $P_{\epsilon}=\sum_{i=1}^3(\sum_kC_{ik}\cdot\sum_{k'}C_{k'i})$.} coefficient \citep{Cohen1960}, which is a measure of the agreement between two raters, in our case the actual and the predicted data. The $\kappa$ coefficient is defined in such a way that the assignment of an object to a certain class due to chance is factored out. The results obtained by these two measures, the $G-mean$ and the   Cohen's $\kappa$ coefficient, are shown in the last two columns of Table \ref{t:methods}, respectively. In both cases, the Random Forest algorithm achieves the best assessment, while there is no clear ranking among the other methods. 

In view of the results presented in Table \ref{t:methods}, we conclude that the best global performance is obtained by the Random Forest algorithm, achieving the maximum percentage of identified stars with the minimum contamination.

\begin{figure*}
  \centerline{\includegraphics[trim=15mm 35mm 15mm 42mm,clip,width=1.4\columnwidth]{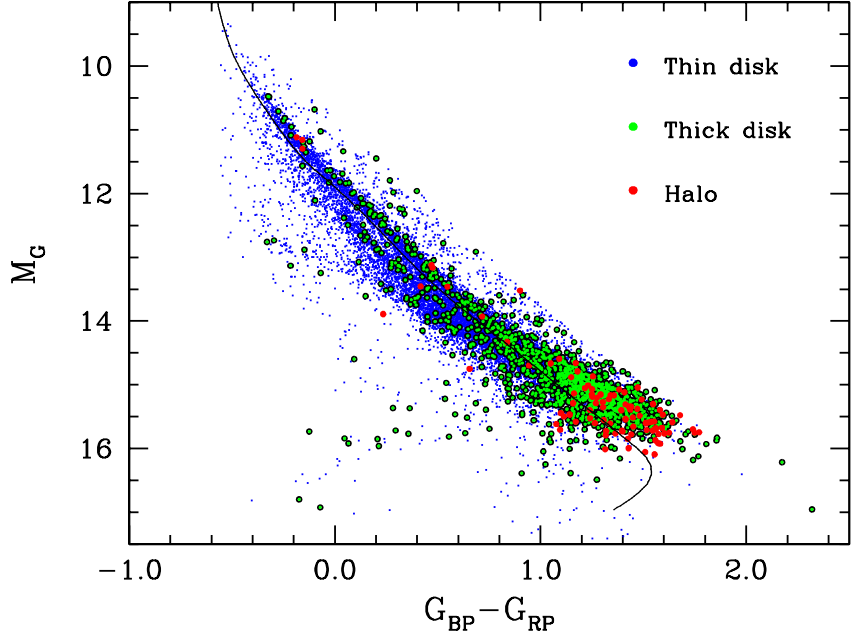}}
  \caption{Color-magnitude diagram for the 100 pc {\it Gaia} sample as resulting after applying our Random Forest classification algorithm. The three Galactic components have been disentangled, obtaining 12\,227, 1410 and 95 white dwarf candidates for the thin and thick disk, and halo population, respectively.} 
  \label{f:hrrf}
\end{figure*}

\section{Classifying the {\it Gaia} 100 pc white dwarf sample.}

In this section we present the result of our classification algorithm applied to our sample of 13\,732 white dwarfs within 100\,pc. We discuss the main characteristics of the Galactic populations identified and derive their most important properties. Specific studies of each population or individual objects are left to forthcoming publications.

\subsection{General properties}

When applied to the observed {\it Gaia} 100\,pc white dwarf sample, our Random Forest algorithm finds 12\,227 thin disk, 1410 thick disk and 95 halo white dwarf candidates\footnote{The complete catalog with their corresponding Galactic component classification can be consulted at: 
http://svo2.cab.inta-csic.es/vocats/v2/wdw}. These values represent a proportion of $89\%$, $10\%$ and $1\%$ for the respective components. However, these are {\sl a posteriori} percentages, that is, after applying our classification algorithm. From the information contained in the confusion matrix (see equation \ref{crf}), we can inverse the problem and then  evaluate the {\sl a priori} percentages. The resulting estimated ratio is then 74:25:1. Thus, the  numerical spatial density is $(3.6\pm0.4)\times10^{-3}\,{\rm pc^{-3}}$, $(1.2\pm0.4)\times10^{-3}\,{\rm pc^{-3}}$ and $(4.8\pm0.4)\times10^{-5}\,{\rm pc^{-3}}$ for the thin disk, thick disk and halo components, respectively. The total density for the 100\,pc disk white dwarf sample found here, i.e. $(4.8\pm0.4)\times10^{-3}\,{\rm pc^{-3}}$,  is in agreement with the recent estimate of the white dwarf space density for the local 20\,pc complete sample by \citet{Hollands2018},  $(4.5\pm0.4)\times10^{-3}\,{\rm pc^{-3}}$, and with the estimate by \cite{Jimenez2018} for a 100 pc sample, $(4.9\pm0.4)\times10^{-3}\,{\rm pc^{-3}}$. Analogously, the space density for halo white dwarfs found here is slightly higher, but compatible with other estimates for the spheroid population (e.g \citealt{Harris2006} found a space density of $4\times10^{-5}\,{\rm pc^{-3}}$). 

The corresponding location of the white dwarfs belonging to the three Galactic components in the color-magnitude diagram is shown in Figure \ref{f:hrrf}, where one can clearly see that both thick disk and, specially, halo white dwarfs concentrate towards the fainter absolute magnitude bins. This fact is in agreement with our expectation of these two Galactic components being formed mainly by old and cool white dwarfs. In particular, for absolute $G$ magnitudes in the range $15<M_{G}<16$, the thin disk contribution is reduced to $65\%$, while the thick disk and halo percentages increase to $32\%$ and $3\%$, respectively. This fact shall be of capital relevance when analyzing the cut-off of the white dwarf luminosity function of the disk, given that at least $35\%$ of the objects at these magnitude bins appear as contaminants from the thick disk and halo populations,  in agreement with previous analysis \citep{Reid2005,Kilic2017}.

\begin{figure}
  \centerline{\includegraphics[trim=5mm 30mm 5mm 35mm,clip,width=1.1\columnwidth]{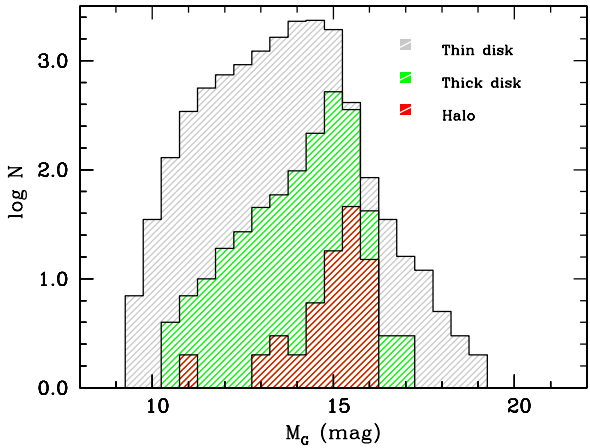}}
  \caption{$M_{\rm G}$ absolute magnitude distributions for our identified populations of the thin disk, thick disk and halo white dwarfs (gray, green and red  shaded histograms, respectively).} 
  \label{f:histog}
\end{figure}

Finally, we analyze the absolute magnitude $M_G$ distributions for the three identified Galactic components, shown in Figure \ref{f:histog}. These distributions can be considered as first approximations to the white dwarf luminosity functions of the three Galactic components. We point out that proper luminosity functions require bolometric corrections, which in turn depend on the individual mass, effective temperature and atmospheric composition of each object. Apart from the typical constant increasing slope clearly visible in each of the white dwarf luminosity functions and the sharp cut-offs, Fig.\,\ref{f:histog} also reveals an extended tail for faint thin disk objects. These faint objects may be associated to ONe-core white dwarfs, given their fast cooling evolution. It has to be emphasised however that a detailed spectroscopic analysis is required to eliminate possible contaminants for these dim magnitudes.

\subsection{Spatial distribution}

\begin{figure}
  \centerline{\includegraphics[trim=10mm 7mm 0mm 65mm,clip,width=0.95\columnwidth]{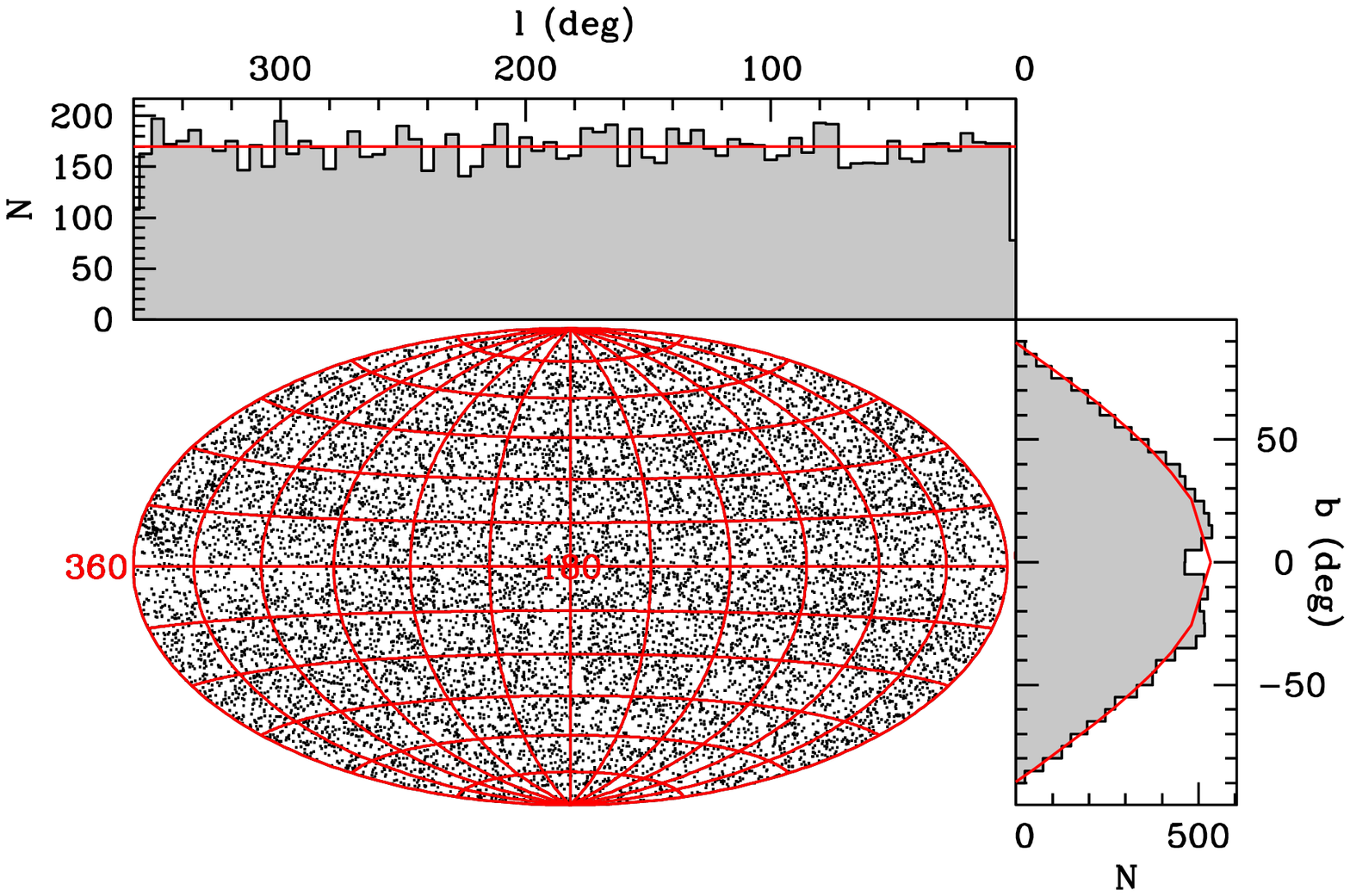}}
   \centerline{\includegraphics[trim=10mm 7mm 0mm 65mm,clip,width=0.95\columnwidth]{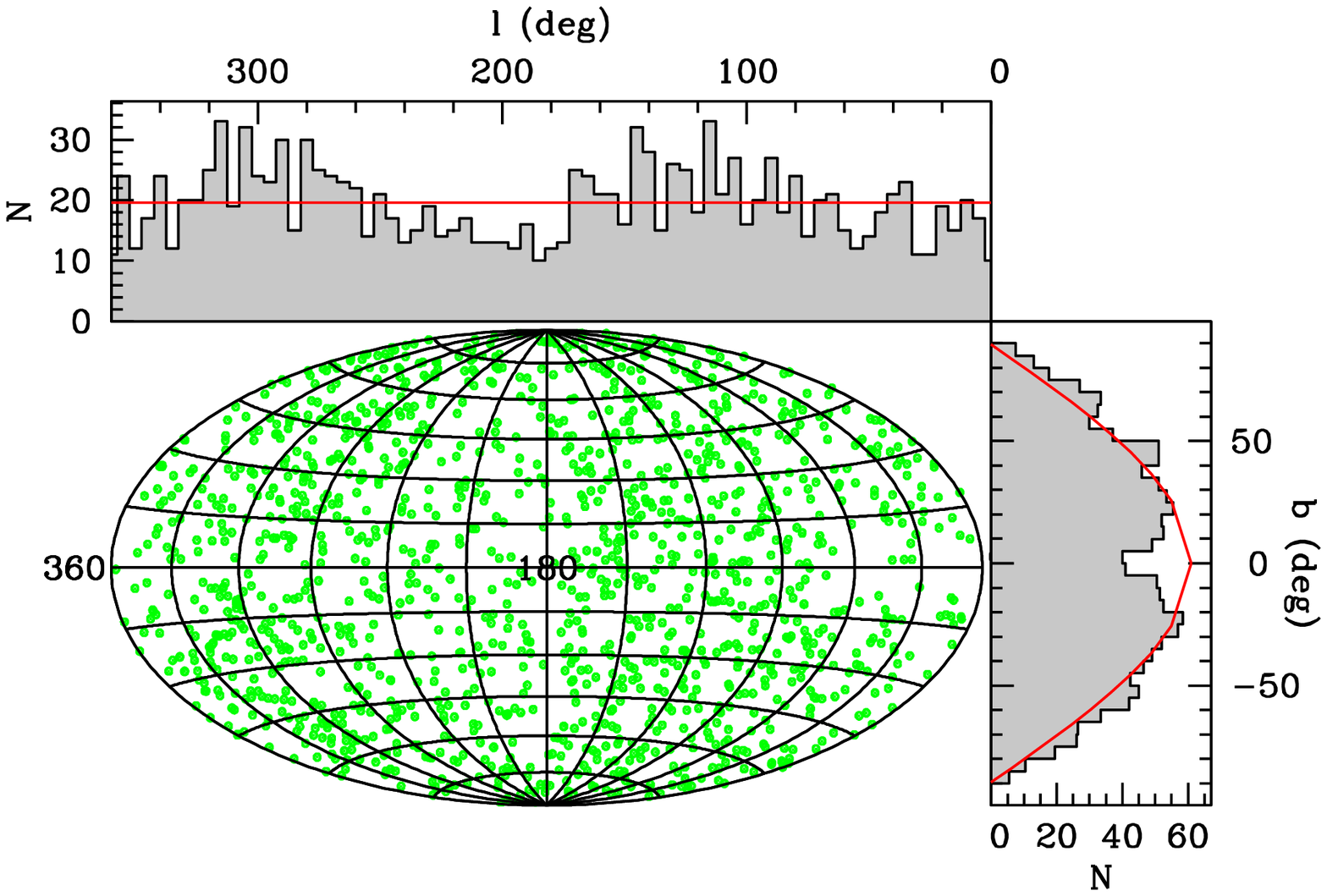}}
    \centerline{\includegraphics[trim=10mm 6mm 0mm 65mm,clip,width=0.95\columnwidth]{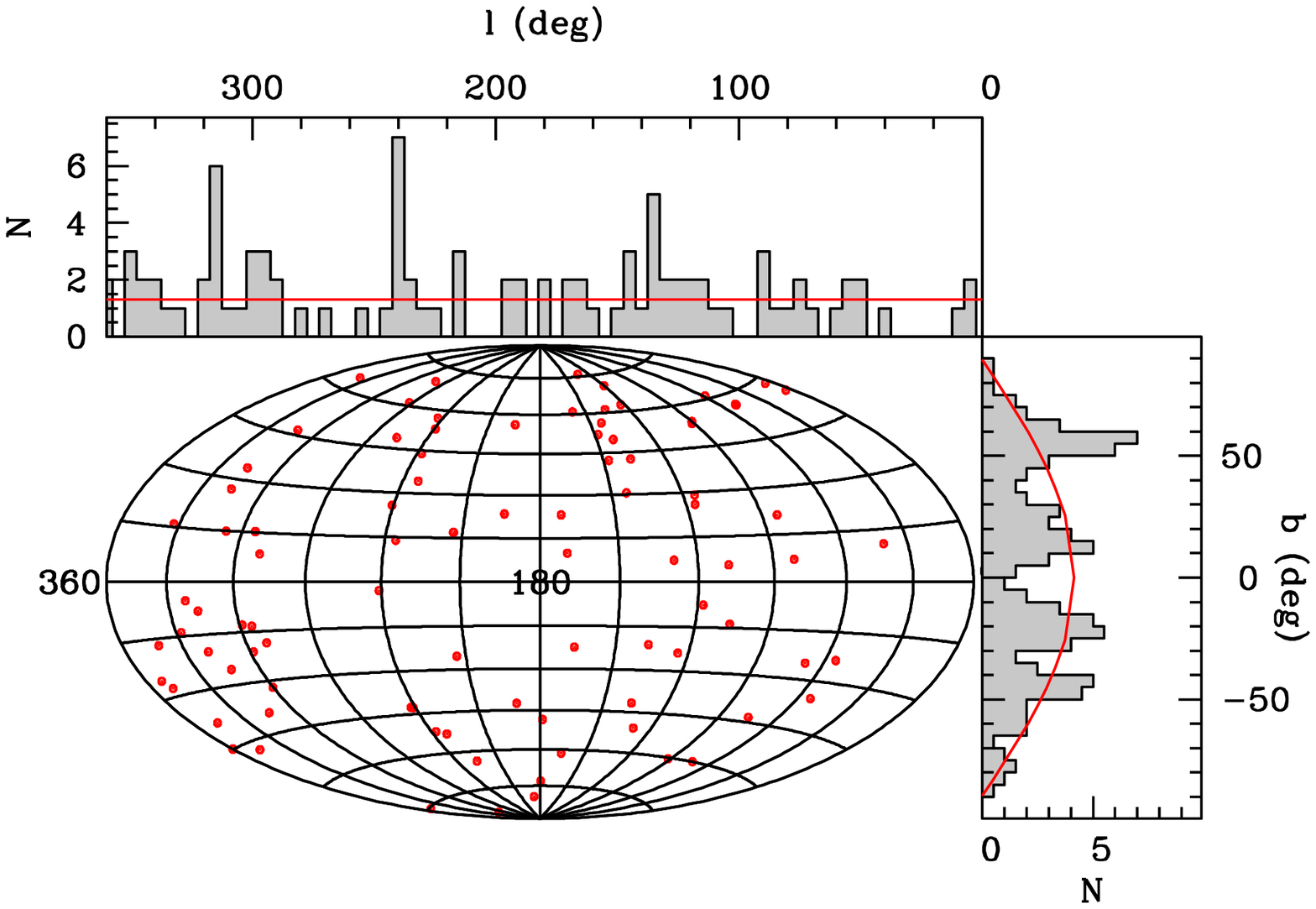}}
  \caption{Aitoff projection  and their corresponding histogram distribution in Galactic coordinates  of the identified white dwarf populations (thin disk, gray dots; thick disk, green dots; halo, red dots).  For visual comparison, a uniform spatial distribution is marked as a red line.} 
  \label{f:aitoff}
\end{figure}

The Aitoff projection in Galactic coordinates for each of the identified white dwarf populations is shown in Figure \ref{f:aitoff} (thin disk, top panel gray dots; thick disk, middle panel green dots; halo, bottom panel red dots). Additionally, we also show the corresponding Galactic coordinate distributions and, for visual comparison, a uniform distribution is marked as a red line. Apparently, all Galactic components present a uniform spatial distribution within $100\,$pc, thus discarding any particular inhomogeneity in the solar neighborhood. A closer look at the thin  and the thick disk populations reveal a slight lack of objects closer to the Galactic plane ($\sim10\%$ and $\sim30\%$, respectively, in the range $-5^{\circ}<b<5^{\circ}$), likely caused by the large number of contaminant objects coming from these crowded regions of the Galaxy, which interfere in our selection process. Concerning the halo population,  additionally to this lack of stars close to the Galactic plane, there seems to appear an excess of objects at  high latitudes, specially in the northern Galactic hemisphere. However, the relative low number of these stars prevents us from achieving any conclusive result.

\begin{figure}
  \centerline{\includegraphics[trim=5mm 30mm 5mm 35mm,clip,width=1.1\columnwidth]{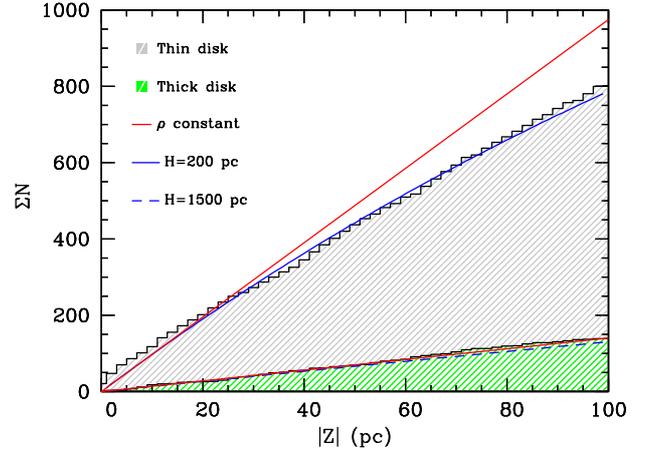}}
  \caption{Cumulative distribution of objects along the coordinate perpendicular to the Galactic plane $Z$ within a local 20 pc radius cylindrical column for the thin disk (gray shaded histogram) and the thick disk (green shaded histogram) white dwarf populations. The thin disk distribution seems to follow a scale height law with $H=200\,$pc (blue solid line). A constant density distribution is also shown (red line). See text for details.} 
  \label{f:hscale}
\end{figure}

We also perform a more detailed analysis of the spatial distribution of the Galactic components regarding the distribution along the coordinate $Z$, i.e.  perpendicular to the Galactic plane. To that end we considered a local cylindrical column centered on the Sun and with a radius in the Galactic plane of 20\,pc. This permits to evaluate the $Z$ distribution up to nearly 100\,pc in height and practically avoiding any type of boundary bias. In Figure \ref{f:hscale} we show the cumulative distribution of objects as a function of the absolute value of the $Z$ coordinate  within our local cylindrical column for the thin disk (gray shaded histogram) and the thick disk (green shaded histogram) white dwarf populations. The halo distribution has been discarded given the poor statistical significance of the sample. For comparative purposes we also plot the cumulative distributions when a constant density of objects is considered  (red line) for the thin and thick distributions. The constant density distribution for the thin disk population has been normalized to the 20\,pc density value given by \citet{Hollands2018}. As it can be seen from Fig. \ref{f:hscale}, the cumulative distribution of our thin disk candidates clearly departs from the constant density model.  Conversely, assuming an exponential decreasing profile (i.e. a scale height profile of H=$200\pm10$\,pc), we suitably fit the observed (thin disk) distribution  (blue solid  line of Fig. \ref{f:hscale}). On the other hand, the thick disk distribution resulting from our selected candidates is compatible with a constant density distribution, or at least with a scale height larger than H\,$>1500$\,pc.

\subsection{Kinematic properties}

\begin{figure}
  \centerline{\includegraphics[trim=5mm 30mm 5mm 35mm,clip,width=1.1\columnwidth]{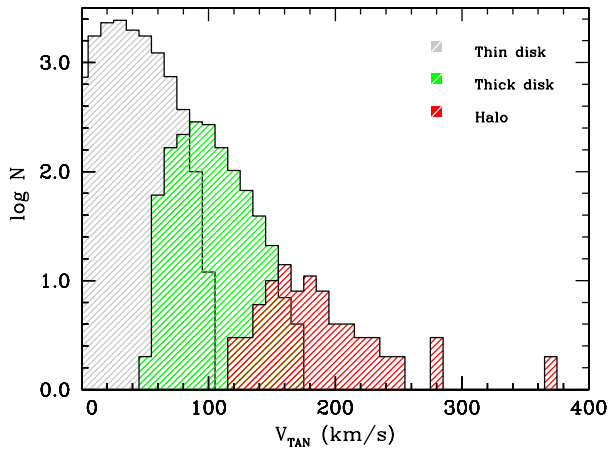}}
  \caption{Tangential velocity distribution for thin disk (gray histogram), thick disk (green histogram) and halo (red histogram) white dwarfs. Our Random Forest algorithm allows unraveling the different Galactic components even if there is an overlap between the distribution. The Figure has been truncated to 400\kms.} 
  \label{f:vtan}
\end{figure}

The corresponding tangential velocity distributions for the thin and thick disk, and halo populations, obtained by our algorithm are plotted in Figure \ref{f:vtan}, clearly showing the overlapping tails of the different distributions. Despite this overlap, our Random Forest algorithm has been able to disentangle the different Galactic components. High-speed thin disk objects can be as fast as average thick disk white dwarfs and, similarly, the fastest  thick disk white dwarfs can have similar tangential velocity values as halo objects. 

Our algorithm is able to clearly identify halo white dwarf candidates as members of a high-speed population, while the thick disk sample presents intermediate kinematic characteristics between the  halo and thin disk populations. It is also worth mentioning here that disentangling the different Galactic components would have not been possible using any other  standard kinematic criterion (see Section \ref{s:compa}). Hence, the overlap in the tangential velocity distributions would remain and, consequently, an inherent and substantial contamination would have stayed in the samples. The use of an 8-dimensional space has permitted our algorithm to maximize the information contained in it, hence achieving a more accurate classification of their components.

\begin{table}
\caption{Mean Galactic velocity values respect to the LSR and their dispersions for the different populations. These values are derived from proper motions and assuming a zero radial velocity component. Hence, the values should be taken as lower limits.}
\label{t:uvw}
\begin{center}
\begin{tabular}{ccccccc}  
\hline \hline 
Component &  $\langle U \rangle $ & $\langle V\rangle$ & $\langle W\rangle$ & $\sigma_U$ & $\sigma_V$ & $\sigma_W$ \\
\hline 
\hline
Thin disk  & 2.52 & -1.45 & 2.60 & 22.5 & 17.4 & 16.2 \\ 
Thick disk & -18.58 & -30.03 & 1.06 & 50.4 & 29.3 & 33.1 \\ 
Halo       & -27.31 & -92.29 & 3.41 & 100.8 & 67.4 & 66.9 \\ 
\hline \hline
\end{tabular}
\end{center}

\end{table}

The corresponding mean values for the Galactic velocity components and their dispersions are shown in Table \ref{t:uvw}. Velocities are relative to the LSR. The peculiar velocity of the Sun  $(U_{\odot},\,V_{\odot},\,W_{\odot})=(7.90,\,11.73,\,7.39)\,$\kms \citep{Bobylev2017} has been adopted. Galactic velocity components have been calculated from proper motions and assuming a zero radial velocity. We recall here that this procedure significantly reduces the mean Galactic velocity components (see Section \ref{s:compa}). Consequently, the mean values presented in Table \ref{t:uvw} should be taken as lower limits. Taking this fact into account, a first glance at Table \ref{t:uvw} reveals that the white dwarf thin disk population is nearly co-moving with the LSR. However, the thick disk population presents a clear velocity lag and an average motion towards the Galactic anti-center. These features are much more accentuated for the halo population, which clearly exhibits a $V-$component decoupled from the disk rotation. Specially unusual are these significant negative $\langle U \rangle$ values, which can not be attributed to the lack of radial velocities, rather than to inhomogeneities in the local sample. A detailed analysis of this issue is left for a forthcoming paper.

\begin{figure}
  \centerline{\includegraphics[trim=5mm 30mm 5mm 35mm,clip,width=1.1\columnwidth]{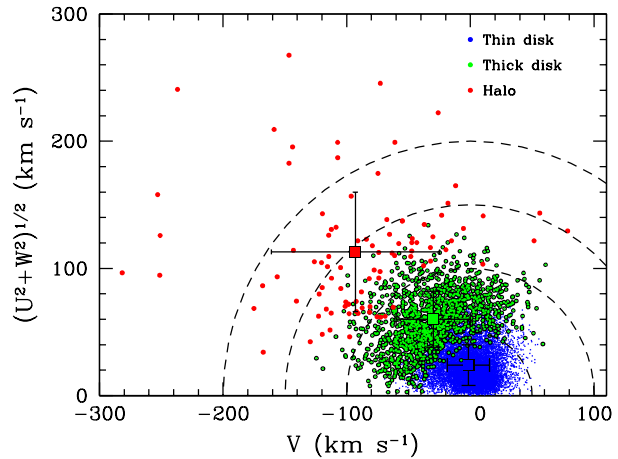}}
  \caption{Toomre diagram for the velocity distribution of thin disk (blue dots), thick disk (green dots) and halo (red dots) white dwarf candidates. Also plotted as big squares their respective centroids with their corresponding $1-\sigma$ error bars. Lines of constant $(U^2+V^2+W^2)^{1/2}$ are shown as dashed lines in steps of $50\,$\kms. } 
  \label{f:toomre}
\end{figure}

In Figure \ref{f:toomre} we show the distribution of the three  identified Galactic components in the Toomre diagram. The colour criteria for the different populations is the same as in our previous figures and lines of constant $(U^2+V^2+W^2)^{1/2}$ are marked as dashed lines in steps of $50\,$\kms. As previously stated, our algorithm permits to identify objects overlapping in the different regions. Objects with speeds below the $150\,$\kms\, line would not  have been classified as halo members if the usual criterion would have been applied. However, our algorithm is able to recover a  significant fraction of low-speed halo members. The identified thick disk population is in agreement with the observed motion for thick F and G dwarf stars \citep[e.g.][]{Bensby2014,Gaiacoll2018}. In a more accentuated way, the overlap exists between the thin and thick disk populations for objects below the $50\,$\kms\, constant line. Additional  information such as the  radial velocities is needed for properly disentangling the different components.

\subsection{Halo white dwarf candidates}

\begin{table*}
\caption{Halo white dwarf candidates within 100 pc identified by our Random Forest algorithm.}
\label{t:halo}
\begin{center}
\begin{tabular}{cccccccccc}  
\hline \hline 
Gaia Source ID & Name & $\mu^*_{\alpha}$ & $\mu_{\delta}$ & $d$  &  $G$ &  $M_{\rm G}$  &  $G_{\rm{\tiny BP}}\!-\!G_{\rm{\tiny RP}}$ &  $V_{\rm TAN}$  &  Ref.  \\ 
 &  & (mas/yr) & (mas/yr) & (pc)  &  (mag) &  (mag)  & (mag) & (\kms)  &    \\ 
\hline 
   420531621029108608  & J001325.85+543807.1  & 776.76  & 600.14  & 32.17  & 18.02  & 15.48  & 1.48  & 149.76  &  \\
  5006232026455470848  & J004521.86-332952.1  & 1826.61  & -1485.17  & 32.92  & 18.64  & 16.05  & 1.51  & 367.58  & $i$ \\
  367799116372410752  & J005516.37+384748.1  & -410.60  & -233.17  & 79.05  & 18.01  & 13.52  & 0.90  & 177.02  & $i$ \\
  2583365245917474816  & J010617.04+114148.5  & 475.28  & -389.01  & 58.53  & 19.62  & 15.78  & 1.44  & 170.48  &  \\
  5042228731477861888  & J012950.76-225714.8  & 394.27  & -200.05  & 84.65  & 20.18  & 15.55  & 1.41  & 177.48  &  \\
  96095735719745280  & J013257.62+194145.6  & 440.72  & 69.65  & 84.54  & 19.52  & 14.88  & 1.15  & 178.89  &  \\
  5142197118950177280  & J014809.05-171231.7  & -105.97  & -1105.83  & 77.19  & 17.56  & 13.12  & 0.47  & 406.68  &  $a,b,i$ \\
  343356212681211392  & J015702.16+393259.6  & 507.13  & -82.71  & 65.14  & 19.73  & 15.66  & 1.45  & 158.74  &  \\
  521409549427243904  & J015856.83+690410.9  & 355.17  & -37.42  & 92.80  & 20.31  & 15.47  & 1.14  & 157.18  &  \\
  2490975272405858048  & J020512.60-051748.1  & 960.22  & 392.98  & 32.36  & 18.14  & 15.59  & 1.53  & 159.23  &  \\
  4616895783694397184  & J023737.80-844521.6  & 759.33  & -57.21  & 58.75  & 19.62  & 15.77  & 1.57  & 212.17  &  $i$\\
  5065611697758431360  & J024813.73-300127.8  & 445.08  & 323.57  & 77.37  & 18.37  & 13.93  & 0.72  & 201.91  & $d,i$ \\
  5188044687948351872  & J030144.18-004446.7  & 107.11  & -550.39  & 69.20  & 19.50  & 15.30  & 1.55  & 184.02  &  $g,h,i$\\
  4862884499360563968  & J034009.29-330105.2  & 492.52  & -334.73  & 54.24  & 19.46  & 15.79  & 1.32  & 153.17  &  \\
  3249657094642979840  & J034213.32-034441.3  & 398.36  & -245.52  & 73.87  & 19.84  & 15.50  & 1.11  & 163.93  &  \\
  4857106909354185344  & J034532.93-361113.2  & 152.66  & -581.56  & 78.53  & 19.85  & 15.38  & 1.47  & 223.93  & $i$ \\
  66837563803594880  & J034647.11+245544.8  & 520.94  & -1157.49  & 39.67  & 18.60  & 15.61  & 1.51  & 238.79  & $h,i$ \\
  4864861112027378944  & J043137.08-381610.9  & 206.49  & -155.74  & 93.07  & 20.59  & 15.75  & 1.77  & 114.15  &  \\
  4864752883148064512  & J043236.94-390202.8  & 700.05  & 721.10  & 33.33  & 17.48  & 14.87  & 1.25  & 158.86  & $i$ \\
  2989049057626796416  & J051850.42-115500.8  & 149.33  & -485.28  & 78.33  & 20.17  & 15.70  & 1.74  & 188.61  &  \\
  192454873200555392  & J055910.46+424838.2  & -63.92  & -519.37  & 94.01  & 18.33  & 13.46  & 0.55  & 233.30  &  $i$\\
  977441274176008192  & J071147.81+460734.7  & -306.67  & -450.93  & 74.64  & 19.15  & 14.79  & 1.18  & 193.04  &  $i$\\
  5613373001468333696  & J073259.02-255825.7  & -79.87  & 366.09  & 99.77  & 20.15  & 15.15  & 1.29  & 177.30  &  \\
  874900643675606912  & J074509.55+262655.3  & 529.69  & -719.19  & 39.55  & 18.75  & 15.76  & 1.60  & 167.54  & $i$ \\
  1110759459929880704  & J074826.81+714156.4  & -67.23  & -376.38  & 70.44  & 19.99  & 15.75  & 1.62  & 127.72  &  \\
  3144837318276010624  & J075014.81+071121.5  & 211.54  & -1782.66  & 18.20  & 16.61  & 15.31  & 1.41  & 154.97  & $i$ \\
  3144837112117580800  & J075015.55+071109.2  & 209.90  & -1790.48  & 18.14  & 16.36  & 15.07  & 1.25  & 155.05  & $i$ \\
  5726927573083821440  & J082219.34-124923.9  & 865.01  & -881.40  & 61.73  & 19.50  & 15.55  & 1.31  & 361.54  & $i$ \\
  5742629217603133056  & J091223.28-095324.0  & -315.36  & 174.01  & 94.89  & 20.17  & 15.29  & 1.16  & 162.09  &  \\
  5215833263797633664  & J091347.45-755302.8  & 175.34  & 480.18  & 65.40  & 19.48  & 15.40  & 1.39  & 158.56  &  \\
  3840846114438361984  & J092531.22+001814.8  & -70.42  & -575.11  & 66.63  & 19.25  & 15.13  & 1.16  & 183.09  & $i$ \\
  1064978578888570496  & J094129.71+651131.9  & 869.35  & -793.43  & 45.62  & 18.46  & 15.17  & 1.36  & 254.62  & $i$ \\
  3836593100382315904  & J100514.02+025416.9  & -704.41  & -4.44  & 83.31  & 19.88  & 15.27  & 1.30  & 278.30  &  $i$\\
  746045096445123968  & J101244.30+323311.9  & 111.49  & -289.18  & 93.44  & 20.57  & 15.72  & 1.40  & 137.33  &  \\
  5192296911732427904  & J103644.11-822557.4  & -485.94  & 116.41  & 60.91  & 19.62  & 15.70  & 1.56  & 144.33  &  \\
  3862858165427681536  & J103654.94+073210.7  & -680.54  & 22.02  & 56.65  & 19.00  & 15.23  & 1.49  & 182.92  &  $i$\\
  1076941716370493696  & J103656.52+711051.6  & -1857.35  & -463.47  & 17.64  & 16.63  & 15.39  & 1.24  & 160.12  &  \\
  855361055035055104  & J104557.38+590428.6  & -1019.19  & -1462.53  & 57.20  & 17.68  & 13.89  & 0.24  & 483.58  & $a,b,i$ \\
  5228861484450843648  & J104957.53-740028.4  & -1022.30  & 601.41  & 42.17  & 19.22  & 16.09  & 1.55  & 237.20  & $i$ \\
  3801499128765222400  & J105356.88-030758.1  & -295.98  & -565.63  & 67.21  & 19.53  & 15.39  & 1.44  & 203.47  & $i$ \\
  3865951435233552896  & J105515.43+081648.7  & -315.37  & -356.20  & 67.14  & 19.61  & 15.48  & 1.68  & 151.48  &  \\
  1055533400343235456  & J110114.86+633345.7  & -139.95  & -577.02  & 82.83  & 19.93  & 15.34  & 1.44  & 233.23  & $i$ \\
  831946229073235200  & J110730.38+485522.1  & -730.18  & -76.89  & 48.07  & 18.61  & 15.20  & 1.27  & 167.36  & $c,g,h$ \\
  5348874243767794304  & J112345.14-515042.7  & -445.38  & 191.95  & 76.80  & 20.03  & 15.61  & 1.34  & 176.63  &  \\
  856513235846126720  & J112353.57+574229.5  & -30.09  & -491.29  & 95.53  & 19.65  & 14.75  & 0.66  & 222.98  &  \\
  5224999346778496128  & J114730.19-745738.2  & -1067.75  & 1354.64  & 19.96  & 17.18  & 15.68  & 1.59  & 163.26  & $i$ \\
  3892524535332945280  & J115131.13+015952.4  & 80.04  & -522.69  & 84.37  & 19.78  & 15.15  & 1.25  & 211.57  & $i$ \\
  5377861317357370240  & J115941.74-463034.2  & -621.47  & 97.71  & 62.84  & 19.18  & 15.19  & 1.25  & 187.49  & $i$ \\
  5377861592235273856  & J115956.86-462903.5  & -621.15  & 98.04  & 63.10  & 17.46  & 13.46  & 0.42  & 188.19  & $i$ \\
  1573358945589364608  & J120514.45+550213.2  & -19.52  & -304.92  & 83.11  & 20.45  & 15.85  & 1.42  & 120.43  & $h$ \\
  3905186270720273152  & J121718.79+083040.0  & -335.18  & -306.25  & 76.18  & 20.14  & 15.73  & 1.32  & 164.02  &  \\
  1533950318546008448  & J123552.20+410937.8  & -510.37  & 189.73  & 60.37  & 19.66  & 15.75  & 1.75  & 155.89  &  \\
  1570514066627694336  & J125007.22+544646.6  & -278.05  & -1253.30  & 23.78  & 17.35  & 15.47  & 1.59  & 144.78  & $i$ \\
  1531097433767946240  & J125506.82+465517.0  & -1092.28  & -113.19  & 37.10  & 18.42  & 15.58  & 1.17  & 193.20  & $c$ \\
  1459546263999675264  & J130316.92+260315.1  & -845.70  & 106.88  & 38.02  & 18.41  & 15.51  & 1.45  & 153.68  &  \\
  6085402414245451520  & J131253.16-472808.8  & -2165.62  & -60.91  & 16.79  & 16.71  & 15.58  & 1.54  & 172.53  &  $i$\\
  3607725941130742528  & J131643.37-153608.6  & -219.55  & -668.78  & 55.93  & 14.86  & 11.12  & -0.19  & 186.70  & $f,i$ \\
  6188655210447329792  & J133800.79-274752.8  & -387.25  & -320.39  & 80.12  & 20.23  & 15.71  & 1.52  & 190.96  &  \\
  6165095738576250624  & J134200.00-341501.8  & -2301.73  & 1147.25  & 20.74  & 16.25  & 14.66  & 1.05  & 252.94  & $i$ \\
  3714266139665215488  & J134807.87+052746.4  & -330.74  & -102.41  & 98.63  & 20.02  & 15.05  & 1.22  & 161.95  &  \\
\hline \hline
\end{tabular}
\end{center}

\end{table*}

\begin{table*}
\contcaption{ }
\begin{center}
\begin{tabular}{cccccccccc}  
\hline \hline 
Gaia Source ID & Name & $\mu^*_{\alpha}$ & $\mu_{\delta}$ & $d$  &  $G$ &  $M_{\rm G}$  &  $G_{\rm{\tiny BP}}\!-\!G_{\rm{\tiny RP}}$ &  $V_{\rm TAN}$  &  Ref.  \\ 
 &   & (mas/yr) & (mas/yr) & (pc)  &  (mag) &  (mag)  & (mag) & (\kms)  &    \\ 
\hline 
  1174809276422844160  & J144206.55+100310.4  & -23.78  & -271.57  & 91.42  & 20.34  & 15.53  & 1.41  & 118.19  &  \\
  1161215296909017728  & J145048.53+073327.0  & -813.59  & -447.93  & 75.04  & 15.53  & 11.16  & -0.16  & 330.49  & $a,b,d,e,i$ \\
  1294793345366747776  & J150002.23+360018.0  & -692.22  & -605.38  & 78.15  & 19.12  & 14.66  & 1.17  & 340.84  & $i$ \\
  1600259390916467072  & J150201.66+540934.9  & -290.61  & -265.88  & 78.98  & 19.83  & 15.34  & 1.43  & 147.54  &  \\
  1612339420228653440  & J150301.99+550942.8  & -901.29  & 70.53  & 74.70  & 20.13  & 15.76  & 1.36  & 320.28  & $i$ \\
  5824436284328653312  & J151745.38-664501.8  & -402.95  & -18.45  & 78.08  & 20.47  & 16.01  & 1.31  & 149.37  &  \\
  6007140379167609984  & J151811.21-380307.2  & -362.82  & -343.80  & 60.99  & 19.61  & 15.68  & 1.56  & 144.57  &  \\
  1277219369981634432  & J152222.97+314600.5  & -127.13  & -298.47  & 85.77  & 20.38  & 15.72  & 1.55  & 131.96  &  \\
  1277232907719022464  & J152340.92+315246.9  & 169.10  & -630.36  & 49.11  & 19.06  & 15.60  & 1.26  & 152.01  &  \\
  5827557213731539328  & J153902.50-612404.7  & -209.01  & -546.29  & 59.87  & 19.11  & 15.23  & 1.33  & 166.06  &  \\
  5817295536128445568  & J170753.04-631915.6  & -327.90  & -506.70  & 53.90  & 18.77  & 15.12  & 1.41  & 154.27  &  \\
  5802598780807649920  & J171529.74-732353.2  & -332.15  & -310.63  & 96.02  & 20.36  & 15.45  & 1.10  & 207.08  & $i$ \\
  1711005951573009792  & J174938.35+824717.5  & -1436.86  & 3314.40  & 16.35  & 14.23  & 13.16  & 0.48  & 280.12  & $a,b,i$ \\
  6363668569344689408  & J181251.72-802824.2  & 100.94  & -287.76  & 99.72  & 20.61  & 15.62  & 1.08  & 144.22  &  \\
  6653858618815379328  & J181454.41-530522.2  & 14.85  & -333.99  & 93.83  & 20.57  & 15.71  & 1.09  & 148.77  &  \\
  4484289866726156160  & J182458.15+121300.1  & -280.65  & -1078.60  & 39.27  & 18.55  & 15.58  & 1.64  & 207.55  & $i$ \\
  2146619161278293248  & J185256.02+533356.2  & -218.00  & 221.60  & 93.90  & 20.60  & 15.73  & 1.40  & 138.43  &  \\
  6663268308043562112  & J192615.74-462738.5  & -59.36  & -526.78  & 86.74  & 19.73  & 15.04  & 1.47  & 218.07  & $i$ \\
  6647162730439433984  & J193631.30-491308.4  & 334.40  & -427.07  & 71.88  & 19.88  & 15.60  & 1.51  & 184.89  &  \\
  2301882675705225472  & J194000.62+834851.5  & -834.34  & -406.05  & 35.79  & 18.29  & 15.52  & 1.24  & 157.50  &  $i$\\
  2082254987541672960  & J200638.18+454451.7  & -507.16  & -204.21  & 66.96  & 19.42  & 15.29  & 1.27  & 173.61  & $i$ \\
  6471523921227261056  & J204235.76-521820.2  & 399.80  & -828.77  & 84.88  & 19.34  & 14.70  & 0.94  & 370.38  & $i$ \\
  1737588947276271744  & J205256.11+070929.5  & -172.08  & -894.23  & 65.70  & 20.00  & 15.91  & 1.57  & 283.74  & $i$ \\
  6580458035746362496  & J211724.53-415650.3  & 516.02  & -303.12  & 58.53  & 19.00  & 15.16  & 1.34  & 166.11  &  \\
  6580551872194787968  & J211724.82-415605.3  & 516.82  & -303.54  & 59.37  & 19.22  & 15.35  & 1.30  & 168.76  &  \\
  1783614400935169408  & J212742.14+154538.2  & -162.27  & -504.91  & 73.91  & 18.67  & 14.33  & 0.84  & 185.89  & $i$ \\
  2687584757658775424  & J212930.26-003417.4  & 1.81  & -426.33  & 97.61  & 20.34  & 15.39  & 1.58  & 197.36  &  \\
  6465689878168451328  & J213946.32-505801.6  & 181.42  & -456.24  & 76.05  & 20.28  & 15.87  & 1.56  & 177.09  &  \\
  2205493129867600256  & J222549.63+635748.2  & 1089.87  & 1344.24  & 38.03  & 17.49  & 14.59  & 1.09  & 312.14  & $i$ \\
  6357629089412187648  & J223035.21-751552.6  & 409.50  & -1827.90  & 15.03  & 16.46  & 15.58  & 1.55  & 133.52  &  \\
  2709539840202060800  & J223707.27+063611.2  & 520.79  & -298.51  & 82.43  & 20.50  & 15.92  & 1.58  & 234.67  & $i$ \\
  1941133391670459648  & J231402.26+454549.7  & -100.99  & -321.38  & 77.56  & 20.44  & 15.99  & 1.43  & 123.90  &  \\
  2631967439437024384  & J231908.88-061314.5  & -605.58  & -1589.82  & 36.21  & 17.87  & 15.07  & 1.38  & 292.18  & $i$ \\
  2641576685735609472  & J234947.69-012454.4  & 470.57  & -324.34  & 64.97  & 19.07  & 15.01  & 1.23  & 176.09  &  \\
  2310942857676734848  & J235418.84-363405.5  & 26.86  & -671.96  & 61.50  & 15.24  & 11.30  & -0.16  & 196.13  & $d,i$ \\
\hline \hline
\end{tabular}
\end{center}
\begin{minipage}{\textwidth}
$^{a}$\cite{Liebert1989}, $^b$\cite{Torres1998} $^c$\cite{Harris2006}, $^d$\cite{Rowell2011}, $^e$\cite{Pauli2006}, $^f$\cite{Kawka2012},  $^g$\cite{Gianninas2015}, $^h$\cite{Si2017},$^{i}$\cite{Kilic2018}
\end{minipage}
\end{table*}

Halo white dwarfs play a key role in our understanding of the Galaxy evolution and has been the focus of interest as a relevant candidate in the dark matter problem, among other important issues (see Section \ref{s-intro}). The lack of a complete and statistically significant sample for these elusive stars has hampered decades of subsequent studies. However, the recent observed  $Gaia$-DR2 provides  a new opportunity in the study of the halo white dwarf population \citep[e.g.][]{Kilic2018}. Giving its relevance we devote this section to specifically analyze our halo white dwarf sample.

Our Random Forest algorithm has identified 95 halo white dwarf candidates within the observed $Gaia$-DR2 $100\,$pc sample. A list containing their {\it Gaia}-DR2 Source identification, Name (in equatorial coordinate format), and some other relevant parameters are provided in Table \ref{t:halo}. Besides, we indicate what objects have been already identified as halo members in previous works. 

Our halo candidate sample has a mean tangential velocity of $\langle V_{\rm tan}\rangle=197\,$\kms, with a minimum value of  $V_{\rm tan}^{\rm min}=114\,$\kms\, and a maximum of $V_{\rm tan}^ {\rm max}=484\,$\kms, consistent with a high-speed kinematic population (see Fig.\,\ref{f:vtan}). Besides, 7 halo candidates are located at a distance closer than 20\,pc. Among these, 4 objects (J1036+7110, J1312--4728, J1749+8247 and J2230--7515) were presented in the analysis of the 20\,pc white dwarf sample of \cite{Sion2009}. It is important to note that in the mentioned analysis these objects were discarded as halo members since  they were not considered old enough, although two of them are actually older than $>8.5\,$Gyr. In a more recent analysis, \cite{Kilic2018} classified two of these objects (J1312--4728 and J1749+8247) as halo members. In particular, J1749+8247 is a low-mass white dwarf, probably formed in a binary system. Even though it has a relatively short cooling time, its progenitor evolutionary timescale is uncertain, hence it may still be consistent with an old population like the Galactic halo.  In any case, these objects deserve further analysis to ascertain their origin.

In their analysis of the Sloan Sky Survey white dwarf luminosity function, \cite{Harris2006} reported 32 white dwarfs with high tangential velocities as halo candidates,  four of them below 100\,pc. Only two of these, J1255+4655 and  J1107+4855, are included in our {\it Gaia} sample and have been classified as halo members by our algorithm. Analogously, \cite{Rowell2011} found 93 white dwarf halo candidates in the SuperCOSMOS Sky Survey, of which only four are located at distances below 100\,pc. These objects are included in our sample, 3 of them (J1450+0733, J0248--3001 and J2354--3634) classified as halo members and one (J1639+8038) belonging to our thick disk population.

Our halo sample also contains three hot objects with $M_{\rm G}<11.5$  classified as a halo white dwarfs. The star J1450+0733 (WD1418+07) was proposed as a halo member due to its high retrograde motion  \citep{Pauli2006} and it is also present in \cite{Rowell2011}. The white dwarf J1316--1536 (NLTT 33503) has a mass estimated $0.51\pm±0.02\,M_{\odot}$ and a likely low-mass progenitor resulting in a lifespan of $\approx 11\,$Gyr \citep{Kawka2012}. Finally, J2354--3634, is also classified as a halo member due to its high velocity \citep{Rowell2011}.

Additionally, we searched for those cool and ultracool white dwarfs presented in the literature as halo candidates and checked whether or not they are included in our list. The two objects reported in \cite{Gianninas2015}, J0301--0044 and J1107+4855, with ages and kinematic characteristics of the halo population, are included in our sample. In \cite{Si2017}, ten halo white dwarfs were proposed, of which only four are in our 100\,pc sample  (the two previous ones also included in \citealt{Gianninas2015} plus J1205+550 and J0346+246), being all four of them identified as halo members by our algorithm. This last object is the well-known and intensively studied ultracool white dwarf WD 0346+246  (J0346+2455) \citep{Hambly1997,Oppenheimer2001,Bergeron2001}. In the same way, the  well known cool white dwarf  F351--50 (J0045-333) discovered by  \citep{Ibata2000} is also classified as halo member by our algorithm.  However, the object LHS 3250 \citep{Harris1999}, mainly due to its slow motion, has been classified as a thin disk member, while the object J1102+4113  \citep{Hall2008}, which have been associated to the halo population due to its extremely low luminosities does not fulfil our photometric excess factor criterion and, therefore, is not present in our {\it Gaia} 100\,pc sample.

Finally, in a recent work, \cite{Kilic2018} found 142 halo white dwarf candidates by using {\it Gaia}-DR2, out of which 55 are within 100\,pc from the Sun. Our algorithm has classified 48 of these objects as halo members. For the remaining seven, four of them (J0544+2602, J0549+2329,  J1240--2317 and J1515+1911) lay out of our limiting region in the HR-diagram, which was specifically defined to mainly select single white dwarfs. The other three (J0034--6849, J0915-2149 and  J1045--1906) have been classified by our algorithm as thick disk objects. Although the kinematic properties of these objects (for instance, their tangential velocities are 160, 161 and $175\,$\kms, respectively) are  compatible with high speed thick disk white dwarfs, the use of only pure-hydrogen or pure-helium atmospheres in our analysis may introduce some error in our photometry assessment. Assuming that the 51 objects from \cite{Kilic2018} and those present in our sample are truly halo members, the three misclassified white dwarfs as thick disk stars would represent only $6\%$ of the sample. This percentage is even smaller than our estimate of $15\%$ (see eq. \ref{crf}) of true halo members classified as thick disk objects, hence reinforcing the idea of robustness of our algorithm. 

It is worth saying that the goal in \cite{Kilic2018} was not to classify the Galactic components but rather to obtain a clean sample of halo white dwarfs. In this sense, they applied a conservative approach by selecting only the targets with velocities that are $5\sigma$ away from the $5\sigma$ distribution of all stars with significant parallaxes. When this criterion is applied to our synthetic population, we verify that the halo sample obtained is practically pure but at the expense of low completeness ($40\%$). The better performance ($80\%$ completeness with a minimum contamination of $5\%$) of the Random Forest as a classification method would explain the larger number of halo white dwarf candidates found in our sample (47 more objects) as compared to the list provided by \cite{Kilic2018}, achiving $\approx80\%$ completness with a minimum contamination $\approx5\%$. As previously stated (see Section \ref{s:compa}), one of the major drawbacks of the Toomre criterion  (and similar ones) is that, when the radial velocities are unknown, the Galactic velocity components are reduced and a significant fraction of low-speed halo members are missed. This fact can be exemplified by the object J1107+4855 which, classified as a halo member not only by us but also by other studies \citep{Harris2006,Gianninas2015,Si2017}, is not included in the \cite{Kilic2018} sample due to its moderately low motion, $V_{\rm tan}=167\,$\kms.

In view of the results presented here, we can conclude that our Random Forest algorithm has succeeded in identifying practically all white dwarfs previously reported as halo members and has significantly extended the sample to those low-speed halo objects that may overlap with thick disk stars. Definitely, our 100\,pc halo white dwarf sample represents the most complete and volume-limited sample up to date.

\section{Conclusions}
\label{s:conclusions}

We have identified 12\,227 thin disk, 1410 thick disk and 95 halo white dwarf candidates belonging to the {\it Gaia} 100\,pc sample by means of an accurate Random Forest algorithm. The unprecedented wealth of valuable information provided to the scientific community by the {\it Gaia}-DR2 and, in particular, the quantity and quality of their data related to the white dwarf population requires the  application of novel artificial intelligence matching learning algorithms in order to extract the maximum information. To this aim we used a supervised Random Forest algorithm to disentangle the different white dwarf Galactic populations given its flexibility and low number of input parameters. The algorithm has been applied to an 8-dimensional space which include astrometric as well as photometric values for each object.  With the aid of a  thorough population synthesis model we accurately reproduce the characteristics of a standard three-component Galactic model. The synthetic population is used for the training of the classification algorithm as well as a testbed for assessing its accuracy. Based on the confusion matrix analysis and on different statistical performance estimators we exhaustively compared our Random Forest algorithm with other classification methods  widely used in the literature. For any of the applied tests, the results show that the performance of the Random Forest algorithm is superior to other classification  methods based on kinematic criteria. Our Random Forest algorithm achieves an accuracy of $85\%$, and it is the method which identified the maximum number of objects with the minimum contamination. Our analysis also demonstrates that those methods based on kinematic criteria, such as the tangential velocity cut, reduced proper motion diagram or Toomre diagram are unable to efficiently resolve the overlapping Galactic populations. On the contrary, the Random Forest algorithm is able to disentangle these populations thanks to the use of the 8-dimensional variable space, resulting all of them relevant in the classification process. 

Once applied to the 100\,pc $Gaia$ DR2 white dwarf sample, the identification of the Galactic populations yields a ratio of 74:25:1 for the thin disk, thick disk and halo white dwarfs, respectively. The corresponding numerical spatial densities are, respectively, $(3.6\pm0.4)\times10^{-3}\,{\rm pc^{-3}}$, $(1.2\pm0.4)\times10^{-3}\,{\rm pc^{-3}}$ and $(4.8\pm0.4)\times10^{-5}\,{\rm pc^{-3}}$. A first study of the main properties of the different identified Galactic population has been done and are summarized as follows: 

\begin{itemize}

\item Concerning the  thin disk population, the performance of our Random Forest algorithm applied to the simulated sample shows that $97\%$ of the white dwarfs are correctly identified and that the contamination (mostly from misclassified thick disk objects) is $13\%$.  The spatial distribution of the $Gaia$ DR2 classified thin disk population follows a clear scale height profile of $H=200\pm10\,$pc, while its motion distribution appears to be co-moving with the LSR. The thin disk population presents a clear cut-off at $M_{G}=15$\,mag, although we estimate a contamination of $35\%$ in the range  $15<M_{G}<16$ of non-thin disk stars. We also identify an extended tail at lower magnitudes, possibly formed by ONe-core white dwarfs. 

\item Although our Random Forest algorithm achieves the highest identification percentage for the thick disk population in the simulated sample ($33\%$), this Galactic component remains as the hardest group to be identified.  The existing overlap in spatial and kinematic distributions and also in photometric properties between the disk populations requires radial velocity measurements to disentangle this population. Nevertheless, the $Gaia$ DR2 thick disk population identified here contains older and cooler white dwarfs than those of a thin disk origin and also reveals a clear lag velocity distribution of $\approx30\,$\kms\ with respect to the LSR. Moreover, the spatial distribution perpendicular to the Galactic plane is compatible with an uniform distribution (or at least with an scale height law with $H>1500\,$pc) for the local sample within 100\,pc. 

\item Finally, a major effort has been put in the analysis of the $Gaia$ DR2 halo white dwarf population. The results show that the capability of our Random Forest algorithm to identify halo white dwarfs is $80\%$ with a very low contamination of $5\%$. We thoroughly checked the literature for already published halo white dwarf candidates below 100\,pc and found that practically all of them are also properly identified by our algorithm. The halo population thus obtained is in accordance with an old and high-speed population clearly decoupled from the Galactic disk rotation. The 95 objects identified by our algorithm constitute the most complete and largest volume-limited sample of the halo white dwarf population to date. 

\end{itemize}

It is expected that forthcoming Gaia Data Releases as well as next-generation wide field surveys will contribute radial  velocity estimates among other relevant parameters that may complement our study. In any case, the identification of the halo white dwarf population, along with the thin and thick disk populations, represents a necessary step towards a better understanding of the star formation history, structure and evolution of our Galaxy.

\section*{Acknowledgements}

This work  was partially supported by the MINECO grant AYA\-2017-86274-P and the Ram\'on y Cajal programme RYC-2016-20254 and by the AGAUR (SGR-661/2017). This work has made use of data from the European Space Agency (ESA) mission {\it Gaia} (\url{https://www.cosmos.esa.int/gaia}), processed by the {\it Gaia} Data Processing and Analysis Consortium (DPAC, \url{https://www.cosmos.esa.int/web/gaia/dpac/consortium}). Funding for the DPAC has been provided by national institutions, in particular the institutions participating in the {\it Gaia} Multilateral Agreement. F.J.E. acknowledges financial support from the Spacetec-CM project (S2013/ICE-2822), and from ASTERICS project (ID:653477, H2020-EU.1.4.1.1. - Developing new world-class research infrastructures). This research has made use of the SIMBAD database,
operated at CDS, Strasbourg, France. We acknowledge the valuable and detailed report of our anonymous referee.

\bibliographystyle{mnras} 
\bibliography{references} 



\bsp	
\label{lastpage}
\end{document}